\documentclass[twocolumn]{aastex63}

\usepackage[utf8]{inputenc}
\usepackage{hyperref}
\usepackage{silence}
\usepackage{graphicx}
\usepackage[caption=false]{subfig}

\usepackage{float}
\usepackage{color,soul}

\usepackage{wrapfig}
\usepackage{bm}
\usepackage{amsmath}

\usepackage{fancyhdr}
\usepackage{wasysym}
\usepackage{natbib}
\usepackage{makecell}
\usepackage{todonotes}
\usepackage{multirow}

\begin{document}

\title{Detection of Near-Infrared Water Ice at the Surface of the (pre)Transitional Disk of AB Aur: Informing Icy Grain Abundance, Composition, and Size}

\author[0000-0002-8667-6428]{S. K. Betti}
\affiliation{Department of Astronomy, University of Massachusetts, Amherst, MA 01003, USA}

\author[0000-0002-7821-0695]{K. Follette}
\affiliation{Department of Physics and Astronomy, Amherst College, Amherst, MA 01003, USA}

\author[0000-0001-6822-7664]{S. Jorquera}
\affiliation{Departamento de Astronom\'ia, Universidad de Chile, Camino El Observatorio 1515, Las Condes, Santiago, Chile}

\author[0000-0002-5092-6464]{G. Duchêne}
\affiliation{Department of Astronomy, University of California, Berkeley, CA 94720, USA}
\affiliation{Université Grenoble Alpes, CNRS, IPAG, F-38000 Grenoble, France}

\author[0000-0002-9133-3091]{J. Mazoyer}
\affiliation{LESIA, Observatoire de Paris, Université PSL, CNRS, Sorbonne Université, Université de Paris, 92195 Meudon, France}

\author[0000-0001-5579-5339]{M. Bonnefoy}
\affiliation{Université Grenoble Alpes, CNRS, IPAG, F-38000 Grenoble, France}

\author[0000-0003-4022-8598]{G. Chauvin}
\affiliation{Université Grenoble Alpes, CNRS, IPAG, F-38000 Grenoble, France}

\author[0000-0002-1199-9564]{L. M. P\'erez}
\affiliation{Departamento de Astronom\'ia, Universidad de Chile, Camino El Observatorio 1515, Las Condes, Santiago, Chile}
\affiliation{N\'ucleo Milenio de Formaci\'on Planetaria (NPF), Chile}

\author{A. Boccaletti}
\affiliation{LESIA, Observatoire de Paris, Université PSL, CNRS, Sorbonne Université, Université de Paris, 92195 Meudon, France}

\author[0000-0001-5907-5179]{C. Pinte}
\affiliation{School of Physics \& Astronomy, Monash University, Clayton VIC 3800, Australia}
\affiliation{Université Grenoble Alpes, CNRS, IPAG, F-38000 Grenoble, France}

\author[0000-0001-6654-7859]{A. J. Weinberger}
\affiliation{Department of Terrestrial Magnetism, Carnegie Institution for Science, Washington, DC 20015, USA}

\author{C. Grady}
\affiliation{Eureka Scientific, Oakland, CA 96002, USA}

\author[0000-0002-2167-8246]{L. M. Close}
\affiliation{Steward Observatory, Department of Astronomy, University of Arizona, Tucson, AZ 85721, USA}

\author[0000-0003-3499-2506]{D. Defrère}
\affiliation{Institute of Astronomy, KU Leuven, Celestijnenlaan 200D, 3001, Leuven, Belgium}

\author{E. C. Downey}
\affiliation{Steward Observatory, Department of Astronomy, University of Arizona, Tucson, AZ 85721, USA}

\author{P. M. Hinz}
\affiliation{Steward Observatory, Department of Astronomy, University of Arizona, Tucson, AZ 85721, USA}
\affiliation{Center for Adaptive Optics, UC Santa Cruz, 1156 High St., Santa Cruz, CA 95064, USA}

\author{F. Ménard}
\affiliation{Université Grenoble Alpes, CNRS, IPAG, F-38000 Grenoble, France}

\author[0000-0002-4511-5966]{G. Schneider}
\affiliation{Steward Observatory, Department of Astronomy, University of Arizona, Tucson, AZ 85721, USA}

\author[0000-0001-6098-3924]{A. J. Skemer}
\affiliation{Department of Astronomy and Astrophysics, University of California, Santa Cruz, Santa Cruz, CA 95064, USA}

\author{A. Vaz}
\affiliation{Steward Observatory, Department of Astronomy, University of Arizona, Tucson, AZ 85721, USA}

\correspondingauthor{S. K. Betti}
\email{sbetti@umass.edu}

\begin{abstract}
We present near-infrared Large Binocular Telescope Interferometer LMIRCam imagery of the disk around the Herbig Ae/Be star AB Aurigae.  A comparison of surface brightness at $K_s$ (2.16~$\mu$m), H$_2$O narrowband (3.08~$\mu$m), and $L'$ (3.7~$\mu$m) allows us to probe the presence of icy grains in this (pre)transitional disk environment.  By applying Reference Differential Imaging PSF subtraction, we detect the disk at high signal to noise in all three bands.  We find strong morphological differences between bands, including asymmetries consistent with observed spiral arms within 100~AU in $L'$. An apparent deficit of scattered light at 3.08~$\mu$m relative to bracketing wavelengths ($K_s$ and $L'$) is evocative of ice absorption at the disk surface layer. However, the $\Delta(K_s-\mathrm{H_2O})$ color is consistent with grains with little to no ice ($0-5$\% by mass). The $\Delta(\mathrm{H_2O}-L')$ color, conversely, suggests grains with a much higher ice mass fraction 
($\sim$0.68), and the two colors cannot be reconciled under a single grain population model. Additionally, we find the extremely red $\Delta(K_s-L')$ disk color cannot be reproduced under conventional scattered light modeling with any combination of grain parameters or reasonable local extinction values. We hypothesize that the scattering surfaces at the three wavelengths are not co-located, and optical depth effects result in each wavelength probing the grain population at different disk surface depths. The morphological similarity between $K_s$ and $\mathrm{H_2O}$ suggests their scattering surfaces are near one another, lending credence to the $\Delta(K_s-\mathrm{H_2O})$ disk color constraint of $<5\%$ ice mass fraction for the outermost scattering disk layer. 
\end{abstract}

\section{Introduction} \label{sec:intro}
The structure of the circumstellar environment of a star affects its evolution and the formation of any bound companions.  The formation of substellar objects within disks is highly dependent on location, stellar and circumstellar activity, and the distribution and state of the materials necessary for formation \citep{Terada2007}.  One key to understanding planet/substellar formation within a disk is understanding the role water ice plays in the formation of both terrestrial and giant planets.  

Water ice in circumstellar disks has been detected in various forms, including as vapor with \textit{Spitzer Space Telescope} \citep{Carr2008, Salyk2008, Blevins2016}, and as crystalline water ice at 44 and 62 $\mu$m \citep{Malfait1999, Meeus2001, McClure2012, McClure2015}.  Near-infrared (NIR) absorption from water ice has also been detected \citep{Pontoppidan2005, Terada2007, Aikawa2012} using spectroscopy of edge-on disks, but these methods lack the angular resolution to resolve the spatial distribution of ice. The presence or absence of ice in circumstellar disks, and in particular the location of the snow line, is important to inform the reservoir of volatiles available to build giant planet cores \citep{Drazkowska2017}.  

In 2008, \citeauthor{Inoue2008} proposed that the presence of icy grains in the surface layers of protoplanetary disks might be inferred from a deficit of NIR scattered light at 3.09 $\mu$m. Icy grains should absorb rather than scatter light at this wavelength preferentially, such that disks with an excess of icy grains will be fainter at 3.09 $\mu$m than the surrounding wavelengths.
This method of constraining the ice distribution and spatial properties has been successfully implemented with detections of icy grains on the surface of the transitional disks of HD 142527 \citep{Honda2009} and HD 100546 \citep{Honda2016}.

However, the recent results of \citet{Tazaki2021} suggest that a more sophisticated approach to modeling the 3.09 $\mu$m ice absorption feature results in disks that require a substantially lower abundance of ice-rich grains in order to match observations.  This was due to a relaxation of the assumptions of simple disk geometry and isotropic scattering present in the initial modeling \citep{Inoue2008}.  
\citet{Tazaki2021} reanalyzed the results of \citet{Honda2009} using anisotropic scattering, and their results suggest a much lower ice abundance, as well as the presence of large micron-sized grains on the surface of HD 142527.   

If we want to understand the reservoir of material available for formation of substellar objects beyond the snow line, we first need to be able to accurately measure its location. The disk surface ice line as probed by scattered light, and the midplane ice line as probed by mm emission \citep{Notsu2017, Notsu2018} may not necessarily align, however, the reservoir of icy grains available both radially and vertically throughout the disk is an important constraint for constraining planet formation theories, mapping grain growth, and informing disk evolutionary processes such as vertical transport. The so-called ``grain filtration" theories are also of particular relevance, as these models predict that small grains, particularly at the disk surface, can easily flow inward past a forming planet while larger grains are impeded by a positive pressure gradient at the edge of the gap formed by a planet and remain there \citep{Zhu2012}.   

AB Aurigae (AB Aur) is one of the most widely-studied Herbig Ae/Be stars.  This A0 spectral type star is located approximately 162.9 $\pm$ 1.5 pc away \citep{GAIA2018} with an estimated mass of 2.4 $\pm$ 0.2 $M_\odot$, and an age of 4 $\pm$ 1 Myr \citep{vandenAncker1997, DeWarf2003}.  It hosts a large, flared, inclined ($i\sim23.2^\circ$) circumstellar disk with a range of substructures.  The extended ($r\sim450$ AU) and massive ($\sim$10 $\mathrm{M_J}$) disk around AB Aur was originally inferred from millimeter observations  \citep{Henning1998}.  Both  spiral structure in the outer disk \citep[$r \gtrsim$ 200 AU,][]{Grady1999, Fukagawa2004, Perrin2009, Tang2012} and a ring-like gap between 40 and 100 AU \citep{Hashimoto2011} have been detected in scattered light.  More recent work using resolved sub-mm and NIR imaging of AB Aur has led to the discovery of inner spiral arms \citep[][Joquera et al. in prep]{Hashimoto2011, Tang2017}, including a twist in the spiral at 30 AU that could be a result of a planet \citep{Boccaletti2020}.  

Furthermore, millimeter continuum and CO gas emission reveal a radial cavity in the disk at 70 AU \citep{Pietu2005}, as well as midplane CO spirals in the inner disk, which could be a result of two or more planets at 30 and $60-80$ AU from the star \citep{Tang2017}.  AB Aur is also accreting at a high rate of $\sim 10^{-7}\ M_\odot \ \mathrm{yr}^{-1}$ \citep{Salyk2013}, indicating ongoing radial transport in the disk.  Due to its intrinsic merit as a possible site of ongoing planet formation and its relatively face-on inclination, we targeted AB Aur to search for icy grains within the disk as they relate to potentially on-going planet formation.

In Section~\ref{sec:methods}, we present new LBTI/LMIRCam observations and discuss our data reduction, while in Section~\ref{sec:flux cal} we detail the methods to obtain accurate disk surface brightnesses.  We present the detection of absorption by icy grains in the disk of AB Aur in Section~\ref{sec:results}. We then discuss how this absorption informs our understanding of ice abundance, grain properties, and optical depth effects in Section~\ref{sec:discussion}.  We conclude in Section~\ref{sec:conclusion} by summarizing the main results.

\section{Observations and Data Reduction} \label{sec:methods}
\subsection{Observations}
AB Aur was observed with the L/M-band InfraRed Camera (LMIRCam) using one of the two 8.4 m mirrors of the Large Binocular Telescope Interferometer (LBTI) on 2015 January 04 \citep{Skrutskie2010} in pupil-stabilized mode. 
The target was observed in two narrow band H$_2$O ice filters (Ice1: $\lambda_c =3.05\ \mu$m, $\Delta\lambda=0.16\ \mu$m and Ice2: $\lambda_c= 3.08\ \mu$m, $\Delta\lambda=0.14\ \mu$m), and the broadband $K_s$ filter ($\lambda_c= 2.16\ \mu$m, $\Delta\lambda= 0.32\ \mu$m).  Due to a significant number of ghosts ($n=5$) in the Ice1 band, a decision was made to obtain additional observations in the ghost-free Ice2 band filter.  To allow for the removal and characterization of thermal emission from the background sky and telescope, the star was nodded up/down by 6\farcs3 on the detector every 120 frames.  Each frame had an exposure time of 2 s, resulting in saturation for the very bright ($K = 4.23,\ L=3.24$) central star within 0\farcs15.
 
The total exposure times for each filter (after removing 43 (1\%) images unusable due to AO system effects such as an elongated point spread function (PSF)) were 31.7 min, 64 min, and 32 min in the Ice1, Ice2, and $K_s$ bands, respectively, resulting in 152$^\circ$ of on-sky rotation.  The full width at half maximum (FWHM) averaged around 0\farcs19 for all wavelengths.

\begin{deluxetable*}{lcccccccccc}[tb]
\centering
\tabletypesize{\scriptsize}
\tablecaption{Log of the LMIRCam AB Aur observations
\label{ABAur}
}
\tablehead{\colhead{Date} & \colhead{Time (UT)} &  \colhead{Target} & \colhead{Band} & \colhead{Seeing (\arcsec)} &  \colhead{Air mass}  & \colhead{$\theta\ (^\circ)$} & \colhead{$t_{exp}$ (s)} & \colhead{N$_\mathrm{exp}$} & \colhead{N$_\mathrm{nods}$} & \colhead{Comment}
}
\startdata
2014-02-13 & 02:18$-$03:39 & AB Aur & $L'$ & 0.66-1.05 & 1.01/1.02 & -66/76 & 0.875 & 50 & 23.5 &  \\
2014-02-13 & 04:04$-$04:36 & HD 39925 & $L'$ & 0.73-1.42 & 1.00/1.01 & 62/75 & 0.875 & 50 & 13.5 & PSF reference\\
2014-02-13 & 04:46$-$06:53 & AB Aur & $L'$ & 0.68-1.41 & 1.02/1.63 & 77/70 & 0.875 & 50 & 27 & \\
2014-02-13 & 07:06$-$07:31 & HD 39925 & $L'$ & 1.02-1.83 & 1.35/1.48 & 73/72 & 0.875 & 50 & 11 & PSF reference\\
2015-01-04 & 02:33$-$03:20 & AB Aur & H$_2$O-3.05 $\mu$m & 1.20-1.00 & 1.23/1.13 & -75/-76 & 2 & 120 & 4 & \\
2015-01-04 & 03:21$-$03:48 & HIP 22138 & H$_2$O-3.05 $\mu$m & -- & 1.09/1.05 & -73/-72 & 2 & 120 & 3 & PSF reference\\
2015-01-04 & 03:48$-$04:07 & HIP 22138 & H$_2$O-3.08 $\mu$m & -- & 1.05/1.03 & -72/-69 & 2 & 120 & 2 & PSF reference\\
2015-01-04 & 04:04$-$04:26 & HIP 22138 & $K_s$ & 0.74-1.03 & 1.03/1.01 & -70/-65 & 2 & 120 & 2.5 & PSF reference\\
2015-01-04 & 04:42$-$04:59 & AB Aur & $K_s$ & 0.76-1.19 & 1.01/1.00 & -74/-65 & 2 & 120 & 2 & \\
2015-01-04 & 05:01$-$05:22 & AB Aur & H$_2$O-3.08 $\mu$m & unk-1.19 & 1.00/1.00 & -63/-20 & 2 & 120 & 2 & \\
2015-01-04 & 05:23$-$05:40 & AB Aur & $K_s$ & unk-3.18 & 1.00/1.00 & 2/60 & 2 & 120 & 2 & \\
2015-01-04 & 05:42$-$06:35 & AB Aur & H$_2$O-3.08 $\mu$m & 0.87-0.96 & 1.00/1.04 & 61/77 & 2 & 120 & 6 & \\
2015-01-04 & 06:54$-$07:11 & HIP 24447 & H$_2$O-3.08 $\mu$m & 0.70-0.92 & 1.03/1.06 & 80/80 & 2 & 120 & 2 & PSF reference\\
2015-01-04 & 07:13$-$07:30 & HIP 24447 & $K_s$ & 0.68-0.76 & 1.06/1.09 & 79/90 & 2 & 120 & 2 & PSF reference\\
\hline
\hline
\multicolumn{11}{c}{Unsaturated}\\
\hline
2014-02-13 & 03:39$-$04:04 & AB Aur & $L'$ & 0.89-0.95 & 1.02/1.04 &  76/77 & 1.455 & 50 & 2.5 & ND 0.9\% \\
2014-02-13 & 06:53$-$07:05 & AB Aur & $L'$ & 1.28-1.02 & 1.64/1.67 &  70/69 & 1.455 & 50 & 1 & ND 0.9\% \\
2015-01-04 & 06:36$-$06:38 & AB Aur & H$_2$O-3.08 $\mu$m & 0.99 & 1.04/1.04 & 77/77 & 0.058 & 100 & 1 & \\
2015-01-04 & 06:39$-$06:46 & AB Aur & $K_s$ & 0.70 & 1.04/1.05 & 77/77 & 0.015 & 100 & 1.5 & \\
2015-01-04 & 06:47$-$06:48 & HIP 24447 & $K_s$ & -- & 1.03/1.03 & 80/80 & 0.015 & 100 & 1 & PSF reference\\
2015-01-04 & 06:49$-$06:51 & HIP 24447 & H$_2$O-3.08 $\mu$m & 0.85-0.86 & 1.03/1.03 & 80/80 & 0.058 & 100 & 1 & PSF reference\\
\enddata
\tablecomments{$\theta$ refers to the parallactic angle in degrees, t$_{exp}$ refers to the exposure time per image, N$_{exp}$ is the number of images per nod, N$_\mathrm{nods}$ is the number of up/down nod sequences.} 
\end{deluxetable*}  

Two PSF reference stars were imaged in the same Ice and $K_s$ bands with the same exposure time as AB Aur the same night shortly after the AB Aur observations.  The two reference stars, HIP 22138 (G8III, $K=4.627$, $L=4.582$) and HIP 24447 (K0, $K=4.219$, $L=4.136$) (nearby and at a similar brightness and color as AB Aur), were observed in each band with 2 s exposures.  HIP 22138 was observed for a total of 24 min, 16 min, and 16 min in the Ice1, Ice2, and $K_s$ bands, respectively. HIP 24447 was observed for 16 mins in each of the Ice2 and $K_s$ band and was not observed in Ice1.  For flux calibration/photometry, unsaturated images of AB Aur ($t_{\mathrm{K}_s}=0.015$ s, $t_{\mathrm{Ice2}}=0.058$ s) and HIP 24447 ($t_{\mathrm{K}_s}=0.015$ s, $t_\mathrm{Ice2}=0.058$ s) were collected in Ice2 and $K_s$.  See Table \ref{ABAur} for a summary of the observations.

We also obtained $L'$ band ($\lambda_c= 3.70\ \mu$m, $\Delta\lambda= 0.58\ \mu$m) observations of AB Aur from Jorquera et al. (2022, in press).  These were collected on 2014 February 12 in non-interferometric pupil-stabilized mode at the LBT.  The star was nodded up/down on the LMIRCam detector every 50 co-adds.  Each co-added consisted of 5 frames, each with exposure times of 0.175 s, resulting in 0.875 s per co-add.  This resulted in saturation of the central star within 0\farcs15.  The total exposure time in $L'$ was 133.5 min.  During the observations, AB Aur transited within 2$^\circ$ of the zenith, resulting in an unstable AO loop due to rapid field rotation, resulting in the loss of a number ($n=62$) of left (SX) image frames.  The FWHM averaged around 0\farcs21.  One PSF reference star, HD 39925 (K5, $K=3.899$, $L=3.52$), was imaged at the same exposure time as AB Aur the same night immediately following the AB Aur observation for a total of 64 mins of integration.  For flux calibration and photometry, unsaturated images of AB Aur ($t = 1.455$ s) were collected using a neutral density filter with a 0.9\% transmission (no unsaturated HD 39925 were observed).         

\subsection{Data Reduction} \label{sec2.2:data reduction}

The Ice and $K_s$ band data were reduced using a custom reduction pipeline following the procedure given in \citet{Crepp2018}. The $L'$ band data was re-reduced using the same pipeline for self-consistency between the three bands.  After removing unusable images, bad pixels ($>$ 8$\sigma$) were corrected by replacing their value with the median of the surrounding 8 good pixels.  Due to offsets between detector channels caused by bias drifts (64 columns), we removed striping from the images by subtracting the 3$\sigma$-clipped median of each channel from its columns; row offsets were removed similarly.  Sky background was removed by subtracting the median of the 120 (50 for $L'$) images taken closest in time at the opposite nod position. 
Distortion correction was also applied, as LMIRCam images contain significant optical distortion from camera flexure. 
The distortion is expected to be constant in time \citep{Maire2015}; therefore, we use an established distortion solution for the DX and SX aperture from 8 November 2017 provided by the LBTO team\footnote{https://sites.google.com/a/lbto.org/lbti/data-retrieval-reduction/distortion-correction-and-astrometric-solution}.  
Finally, LMIRCam is designed to be used with both apertures; since AB Aur was imaged on one side in $K_s$ and Ice2, the images are oversampled, and we therefore bin each image into $2\times2$ pixels, giving a final plate scale of 0\arcsec.022.  We treat each eye taken with $L'$ independently in order to match $K_s$ and Ice.  Therefore, we also bin each image into $2\times2$ pixels.

Images are registered across all three bands using cross correlation, and the unsaturated images are median combined. To fit for and remove the effects of the central star before derotating and stacking the frames, we use reference differential imaging (RDI) with Karhunen-Lo\`{e}ve Image Projection \citep[KLIP;][]{Soummer2012} using the \texttt{pyKLIP} python package \citep{Wang2015}.  
KLIP works by creating a model of the stellar PSF for each frame from a library of reference images, which can then be subtracted to remove the stellar PSF.  
In the case of RDI, these reference images are images of one or several `reference' stars, selected to be close in brightness and type, and without disks or planet detected around them.

\begin{figure*}[htp]
\centering
\includegraphics[width=\linewidth]{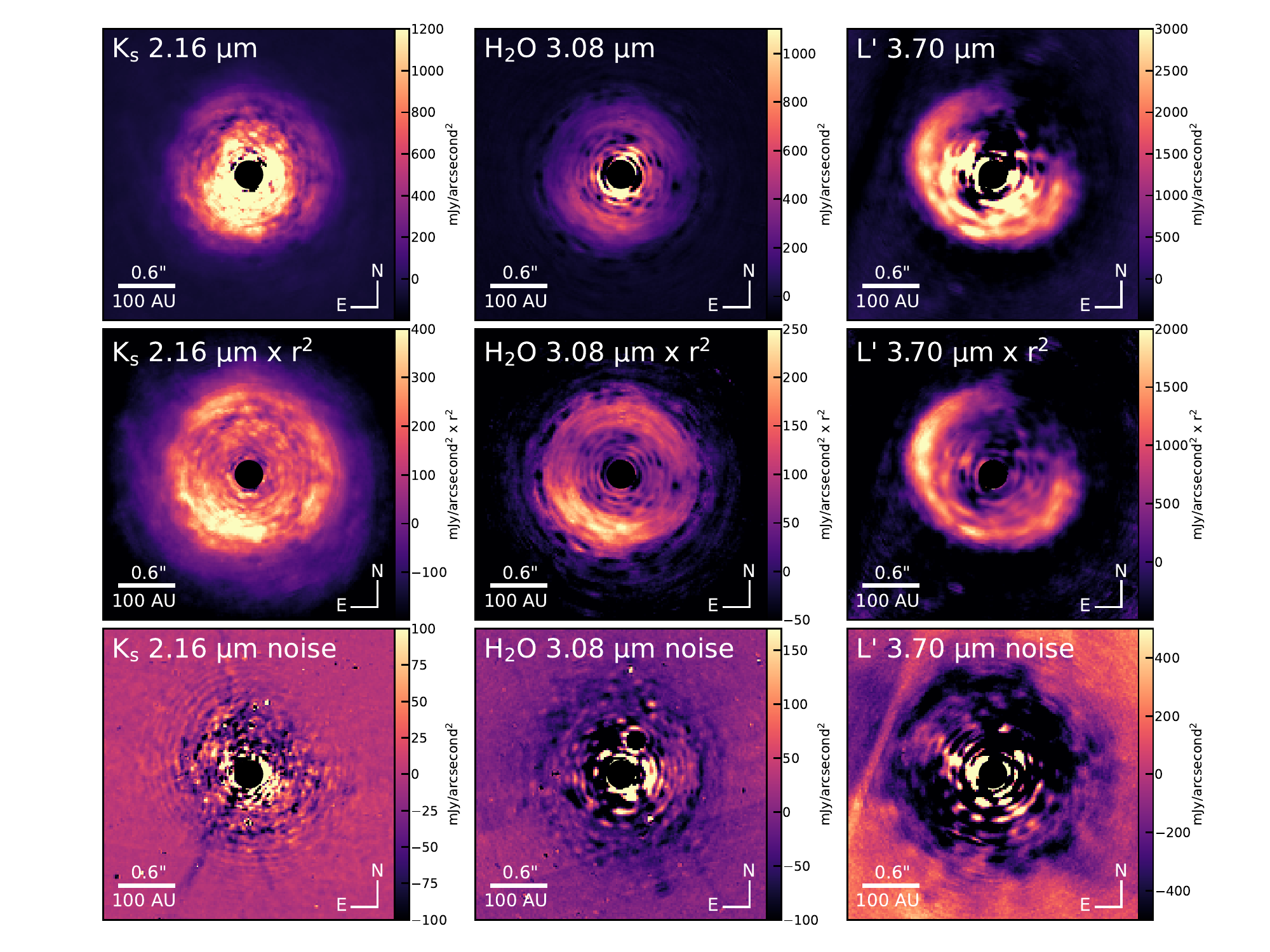}
 \caption{\textbf{top)} KLIP-RDI reduced images of the disk around AB Aur at $K_s$ (left), 3.08 $\mu$m Ice (center), and $L'$ (right). \textbf{middle)} Same at top panel but multiplied by r$^2$ to enhance the outer substructure.  \textbf{bottom)} Uncertainty noise maps from the KLIP-RDI reduction of the PSF reference stars. The field of view is 3\arcsec $\times$ 3\arcsec with north-up and east-left.  A software mask (black) of 0\farcs15 in radius is centered on each image. A black software mask also covers a ghost in the H$_2$O 3.08 $\mu$m noise map NW of the central mask. }
 \label{disks}
\end{figure*}

We select the 100 most correlated reference star images to form the PSF library for each frame, and then use KLIP principal component analysis (PCA) to construct the model PSF and subtract it from the target image.  The number of eigenvectors (KL modes basis vectors) used controls the complexity of the constructed PSF. We apply a simple (3 KL mode) model in order to avoid, to the extent possible, oversubtraction of the nearly symmetrical and nearly face-on disk of AB Aur.
Finally, KLIP derotates and stacks all the AB Aur frames to create a final image (Figure \ref{disks}).

In order to demonstrate its effectiveness and estimate uncertainty, we also applied KLIP-RDI to the reference images, using one PSF reference star as a reference library for the other.  If the disk structure around AB Aur is robust, similar structures should not appear on the PSF-PSF KLIP-RDI images.  As the PSF reference stars are at the same brightness as AB Aur for their observed relevant bands (see Table~\ref{referencestars}), the resulting images give an estimate of the resulting noise due to PSF mismatch (bottom panel of Figure \ref{disks}).  Following \citet{Chen2020}, we determine the uncertainty on recovered flux at each radial location by calculating the standard deviation in concentric annuli around the star with a width equal to the FWHM of the PSF ($\sim4$ pixels). 

\begin{deluxetable}{lcccc}[htp]
\centering
\tabletypesize{\scriptsize}
\tablecaption{Properties of Observed Targets
\label{referencestars}
}
\tablehead{\colhead{} & \colhead{AB Aur} & \colhead{HIP22138} & \colhead{HD 24447} & \colhead{HD 39925}
}
\startdata
R.A. (J2000)* & 04:55:45.84 & 04:45:50.07& 05:14:39.43 & 05:57:00.23\\
Decl. (J2000)* & +30:33:04.29 & +28:39:38.47& +31:24:06.93&+30:36:34.72\\
Observed & all& $K_s$/Ice& $K_s$/Ice& $L'$\\ [-1ex]
\hspace{0.1cm} band & & & & \\
$K$ (mag)$^\times$ & 4.23$\pm$0.02 & 4.62$\pm$0.02& 4.22$\pm$0.02 & 3.89$\pm$0.23\\
$L$ (mag)$^\wedge$ & 3.24$\pm$0.13& 4.58$\pm$0.07& 4.13$\pm$0.10& 3.52$\pm$0.11\\
$K-L$ (mag) & 0.99$\pm$0.14& 0.04$\pm$0.08&0.09$\pm$0.10& 0.37$\pm$0.26\\
\enddata
\tablerefs{*\citet{GAIA2018}, $^\times$\citet{Cutri2003}, $^\wedge$\citet{Cutri2012}} 
\end{deluxetable}

\section{Disk Surface Brightness} \label{sec:flux cal}
In order to quantify flux loss as a result of KLIP-RDI over-subtraction and PSF mismatch, which must be corrected for in order to report accurate photometry, we forward model synthetic MCFOST disks to determine the ``throughput" of the algorithm at each band.  These forward models were selected to provide a good match to the the surface brightness of the disk, as this is what we aim to quantify. We note that we were unable to find MCFOST models that both (a) converged on physically-realistic values of geometric and grain properties and (b) provided a good match to the surface brightness of the disk.  Therefore, we use the models only for correction of surface brightness flux loss from KLIP-RDI, and do not make inferences from them about the physical or grain properties of AB Aur.  See Appendix~\ref{appendix:diskmodel} for full details of our modeling attempts.

\subsection{Throughput Correction}
After computing the best fit models for each wavelength independently, we compute the throughput (TP) of the KLIP-RDI algorithm for these models.  
The TP is computed as the ratio of the initial, convolved model disk to the final KLIP-RDI forward modeled image, with a value of 1 indicating no flux loss, and 0 indicating all disk flux was removed (oversubtracted) by RDI. 

\begin{figure}[htp]
\centering
\includegraphics[width=\linewidth]{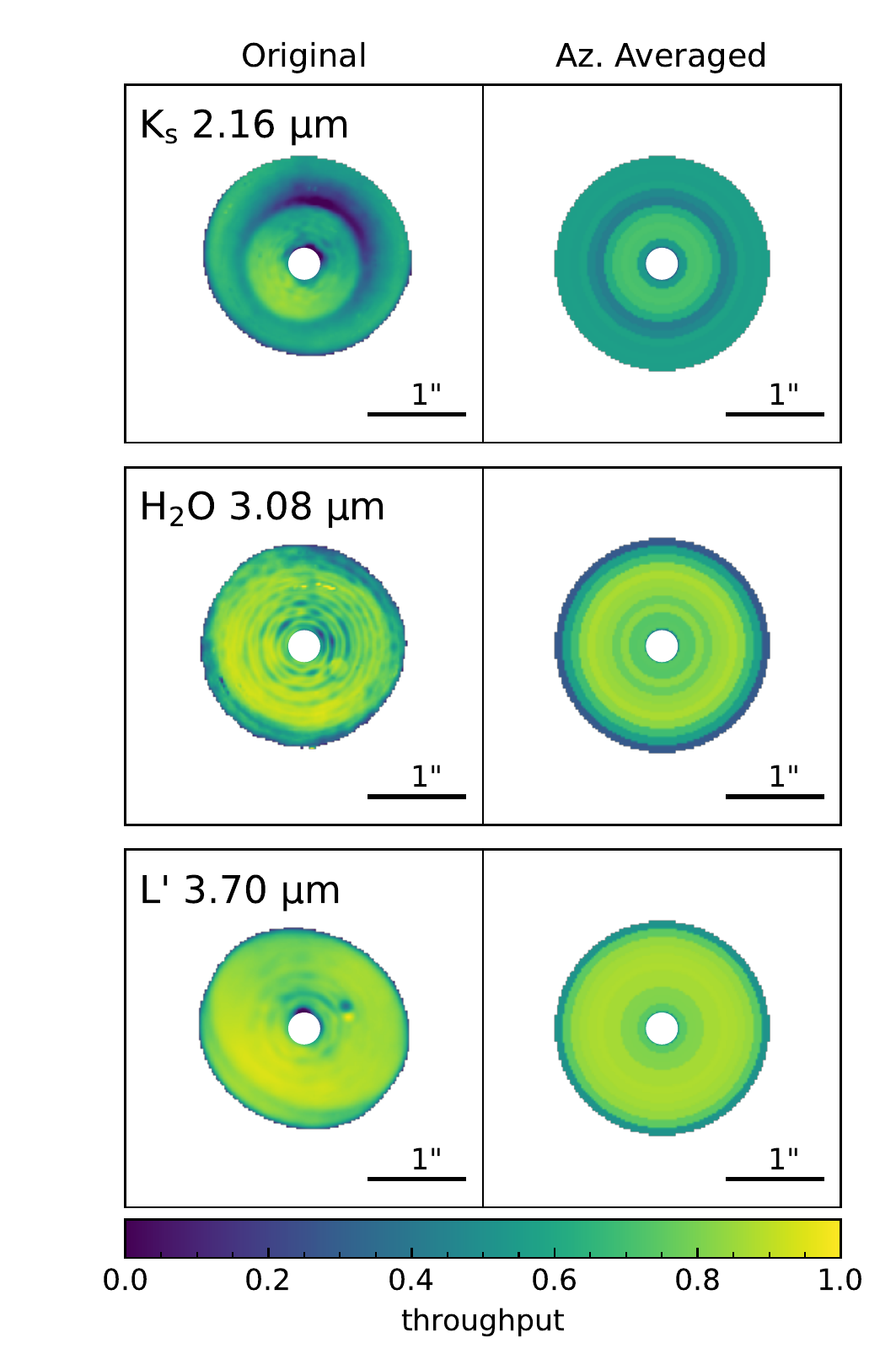}
 \caption{Map of throughput resulting from the KLIP RDI reduction for $K_s$ (top), 3.08 $\mu$m Ice (center), and $L'$ (bottom) filters.  
 The left column shows the true throughput and the right column is the azimuthally averaged throughput.  
 A value of 0 means all flux is removed by RDI, while a value of 1 means no flux is lost.  The ``point source" in the west of the $L'$ TP map is an artifact in the post-processing. }
 \label{throughput}
\end{figure}

In Figure~\ref{throughput}, the left column show the throughput maps computed for the best-fitting MCFOST model.
As the substructure seen in the TP is mainly a result of broadly symmetric speckle noise or post processing artifacts, we azimuthally average\footnote{Boxcar smoothing the TP map to remove the variations gives similar final results/conclusions. We have chosen to azimuthally average the TP to remove any non-axisymmetric variations.} the TP using annuli of width $\sim$FWHM (as shown in the right column of Figure~\ref{throughput}), which smooths over these spurious variations.  
We find a consistent degree of flux loss (TP$\sim15-25\%$) in 3.08 $\mu$m ice and $L'$ bands, and almost 45\% flux loss in $K_s$ band.  At the outer boundary of the disk region, the model flux drops below the speckle noise in the forward model.
The final reduced KLIP-RDI AB Aur images are divided by the final azmithually averaged TPs to correct for algorithmic flux loss before conducting photometry on the disk.
In Appendix~\ref{TPuncertainty}, we show that varying the surface brightness and geometry of the initial model has a negligible effect on the measured throughput ($\lesssim2.5\%$).  

\subsection{Photometric Calibration} \label{photocal}

\begin{figure*}[htp]
    \centering
    \includegraphics[width=\linewidth]{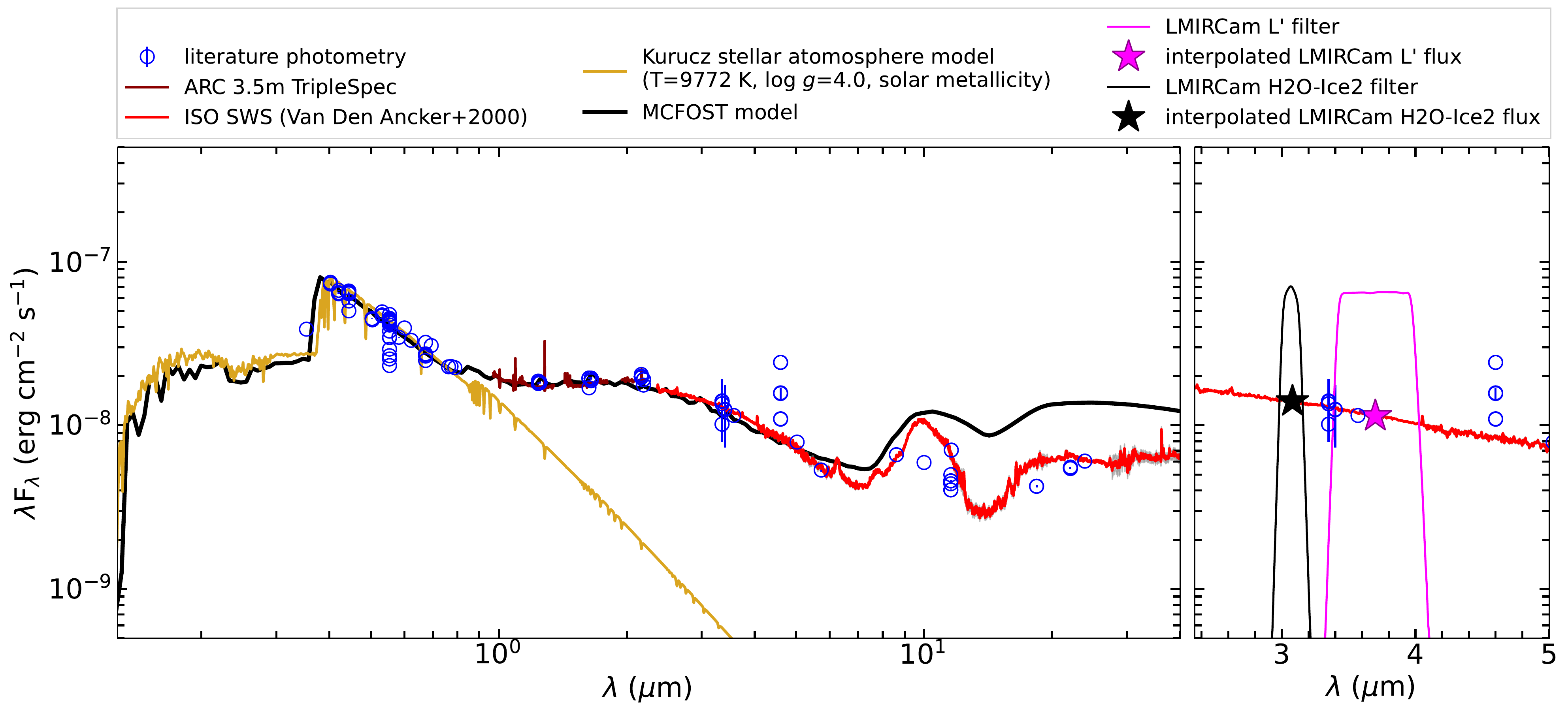}
    \caption{\textbf{left)} optical to FIR SED of AB Aur.  A reddening extinction correction is applied using $A_V = 0.431$ mag \citep{Vioque2018} assuming an \citet{Cardelli1989} extinction law. The blue circles indicate known observed photometry from literature (photometry downloaded from VizieR photometry viewer).  The dark red spectra is from $1-2.46$ $\mu$m taken from ARC 3.5m TripleSpec at APO.  The red spectra is from ISO SWS \citep{vandenAncker2000}.  The gold line shows the best fit Kurucz stellar atmosphere model in the optical (T$=9772$ K, $\log g = 4.0$). The thick solid black line is the best fit MCFOST model.   \textbf{right)} AB Aur spectra from $2.4-5\ \mu$m overlaid with the literature photometry (blue), and the LMIRCam H2O-Ice2 (thin black line) and $L'$ filters (magenta line).  Synthetic photometry for each filter is shown by the corresponding colored star.}
    \label{SED}
\end{figure*}

To determine the surface brightness of the throughput corrected disk, we photometrically calibrate the disk following the procedure of \citet{Quanz2011}.  The SED of AB Aur (Figure~\ref{SED}) shows strong IR excess from 1-4 $\mu$m in both photometry and spectroscopy indicating that the innermost disk component \citep[within 0.3 AU;][]{Tannirkulam2008, Natta2001} is completely unresolved.  Therefore, we can assume that our unsaturated images of AB Aur contain both star and excess inner disk thermal radiation.  We use the unsaturated images of AB Aur to determine both the instrumental zeropoint and the photometric system zeropoint since the known photometry and spectroscopy of AB Aur also contain this excess IR emission.
To quantify the instrumental zeropoint, we use aperture photometry to measure the total intensity of the unsaturated AB Aur images in each band and translate this to an instrumental magnitude.  We find $1.9\times10^{6}$ counts/s, $1.0\times10^{6}$ counts/s, and $9.4\times10^{5}$ counts/s in the $K_s$, 3.08 $\mu$m and $L'$ filters, respectively.  For $K_s$ band, the transmission bandpass is centered at 2MASS $K_s$ band; therefore, we can use the known 2MASS AB Aur $K_s$ magnitude \citep[$K_s$ = 4.23 $\pm$ 0.016; ][]{Cutri2003} to calculate the zeropoint. 
 
We calculate the 3.08 $\mu$m and $L'$ zeropoints by convolving the LMIRCam transmission profiles with the published ISO/SWS spectrum of AB Aur \citep[$2.4-45\ \mu$m][]{vandenAncker2000}. The transmission curves for the filters were obtained from the SVO Filter Profile Service \citep{Rodrigo2012, Rodrigo2020}.  The $2.6-40\ \mu$m flux calibrated spectrum (Figure~\ref{SED} left panel) covers the Ice2 and $L'$ LMIRCam bands, allowing us to synthesize the photometric flux in each band (see right panel Figure~\ref{SED}).  Using the LMIRCam known photometric system zeropoints \citep[3.08 $\mu$m: 359.73 Jy, $L'$: 257.10 Jy; ][]{Rodrigo2012, Rodrigo2020}, we convert flux to magnitudes.  We then calculate and apply the instrumental zeropoints to find the surface brightness of the disk using the photometric system zeropoints ($K_s$:~666.7 Jy, 3.08 $\mu$m and $L'$ as listed above).

\section{Results} \label{sec:results}

\begin{figure}[htp]
    \centering
    \includegraphics[width=\linewidth]{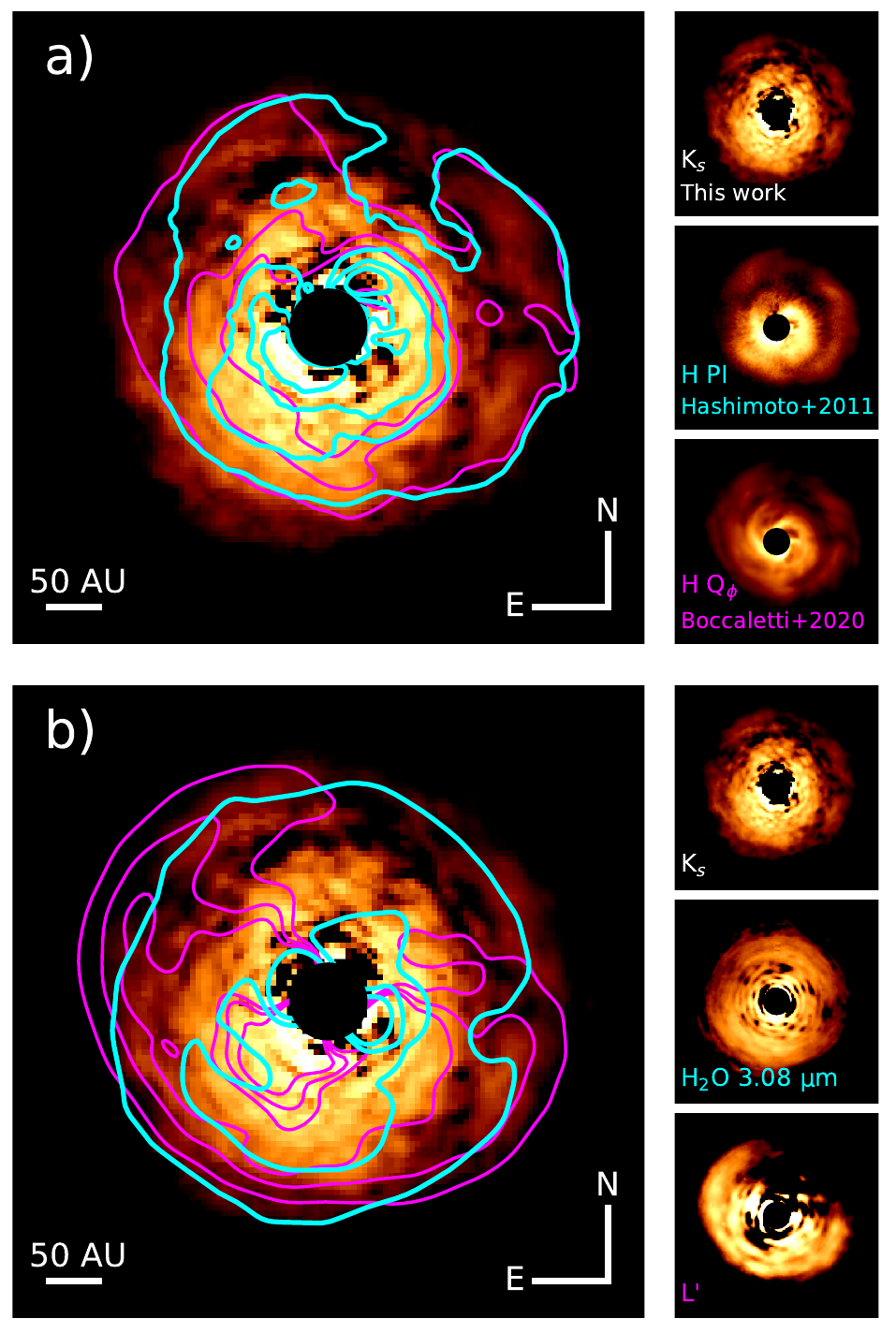}
    \caption{\textbf{a)} \textit{Main panel}: $K_s$ band total intensity imagery from this work overlain with $H$-band polarized intensity contours from \citet{Hashimoto2011} (cyan) and \citet{Boccaletti2020} (magenta). \textit{Right hand inset}: The three individual $K_s$ band total intensity morphology closely matches that of $H$-band PI imagery, for which PSF subtraction is unnecessary, and suggests that KLIP-RDI is able to accurately recover the disk surface brightness morphology. \textbf{b)} \textit{Main panel}: $K_s$ band total intensity imagery overlain with contours of the H$_2$O 3.08 $\mu$m band (cyan), and L' band (magenta). \textit{Right hand inset}: The three individual images are on equivalent scales (top to bottom: $K_s$, 3.08 $\mu$m, and $L'$). These reveal that the observed H$_2$O 3.08 $\mu$m morphology most closely resembles $K_s$ band. This suggests that the scattering surfaces at these two wavelengths are likely at similar depths in the disk surface layers. The poor qualitative and quantitative match of the $L'$ morphology to the other two wavelengths suggests that the $L'$ scattering surface is likely located deeper in the disk surface layer.} 
    \label{contours}
\end{figure}

\begin{figure}[htp]
    \centering
    \includegraphics[width=0.7\linewidth]{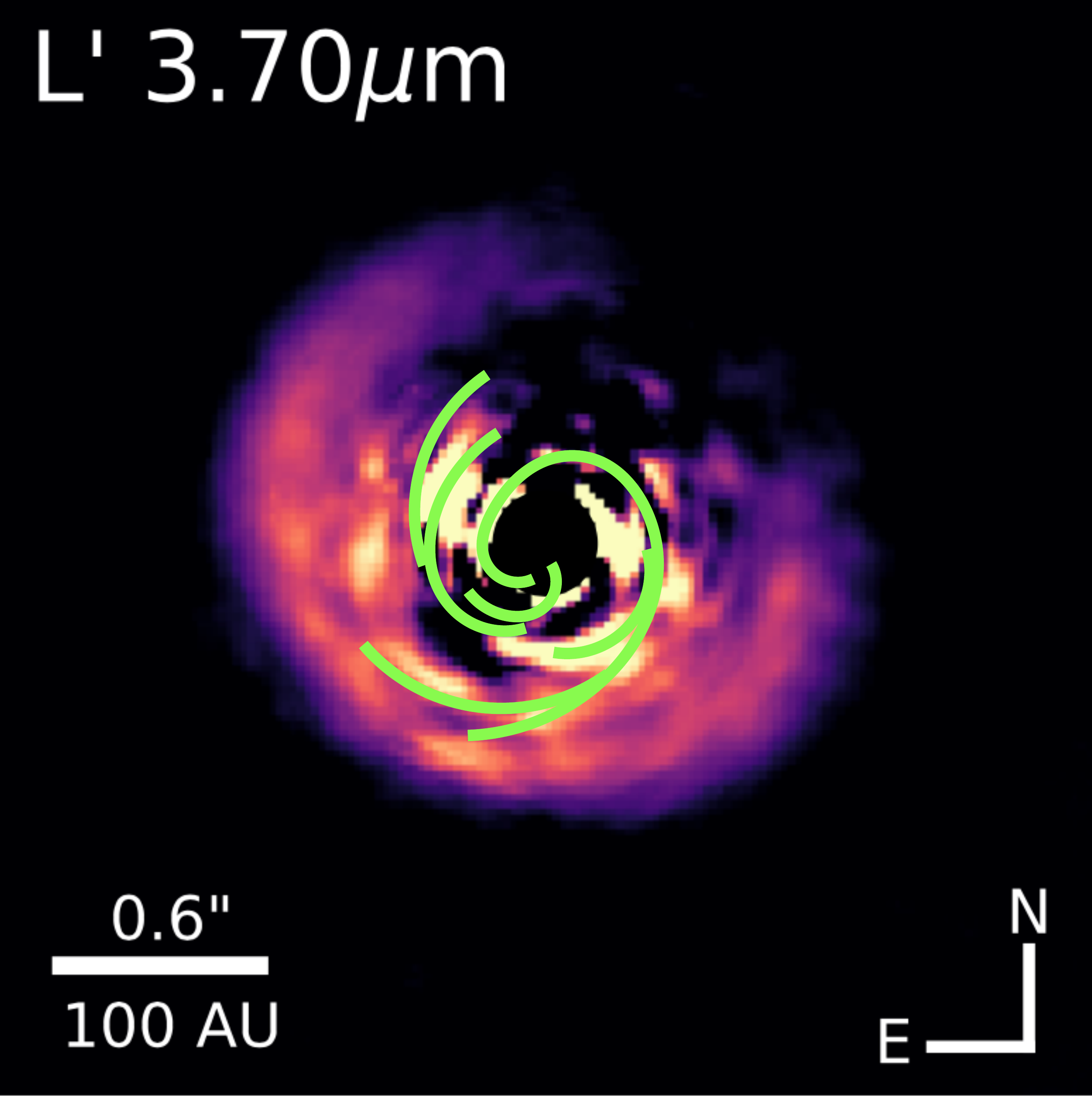}
    \caption{$L'$ KLIP-RDI  reduced image of the disk of AB Aur overlaid with the locations of the spiral arms from \citet{Boccaletti2020} in green.  The spiral arms align with areas of brighter flux in the $L'$ image indicating they are substructure in the disk and not artifacts from data reduction.} 
    \label{overlay}
\end{figure}

We detect the disk of AB Aur with a signal to noise ratio (SNR) of $\sim2.5$ from $\sim$~25 to $\sim$~175 AU in all three bands, as shown in Figure \ref{disks}.  Overall, as shown by Figure~\ref{contours}a, the $K_s$ disk structure and morphology closely match the morphology of the $H$-band polarized intensity images of \citet{Hashimoto2011} and \citet{Boccaletti2020}. As PI imagery is generally considered extremely robust, and is virtually unaffected by stellar emission, this match lends credence to the integrity of our KLIP-RDI total intensity disk recoveries in general, and $K_s$ band in particular. The SE rim of the disk is markedly brighter than the NW in $K_s$ and H$_2$O 3.08 $\mu$m band, similar to the $H$ band polarized images of \citet{Hashimoto2011}, suggesting strong forward-scattering at $K_s$. 
The disk appears more elliptical in $L'$ (see Figure~\ref{contours}b), with a shape and spirals features similar to those reported in Jorquera et al. (2022, in press), \citet{Boccaletti2020}, and \citet{Oppenheimer2008}.  Figure~\ref{overlay} show the locations of known spiral arms and substructure \citep{Boccaletti2020} overlaid on the $L'$ RDI image.  The location of the spiral arms aligns with visible substructure in $L'$, visually confirming that we are seeing the spirals.

\subsection{Surface Brightness Profiles}
Figure~\ref{profiles} shows disk surface brightness profiles along the major (panels a, c, e) and minor (panels b, d, f) axes of the disk in contrast units (panels a and b), flux units (panels c and d) and $K_s-\mathrm{H_2O}$ and $K_s-L'$ disk colors (panels e and f) space.  Locations of apparent negative flux are indicated by dashed lines.  We are able to detect the disk down to $10^{-5}$ contrast.  The substructure in $L'$ band is clearly seen in panels c) and d) as peaks in the radial profile, while the structure in $K_s$ and ice bands matches that of \citet{Hashimoto2011}.   From these profiles, we also see that the Ice band is fainter than $K_s$ and $L'$ at all distances, lending support to absorption by icy grains.
Finally, in panels e) and f), we measure the color excess of the scattered light from the disk (see Section~\ref{sec4.2:spectra} for more details).  We find that the disk is markedly red in $\Delta(K_s - L'$) at all separations, indicating a positive spectral slope. Conversely the disk is blue in $\Delta(K_s - \mathrm{H_2O})$ at most separations, suggesting the presence of water ice in the disk surface layer.  Red $K_s - L'$ disk colors have been associated in the literature with large grains that strongly forward scatter at shorter wavelengths; the grain scattering efficiency at these short wavelengths is lowered causing the longer wavelength light ($L'$) to appear brighter \citep{Mulders2013, Tazaki2019}.  We discuss the strong red disk color in Section~\ref{sec:opticaldepths}. 

\subsection{Icy Grains Across the Disk}
The model predictions of \citet{Honda2016} and \citet{Tazaki2021} suggest that the depth of water ice absorption across the disk can vary even if the ice abundance remains fairly consistent. 
However, we find that the observed spectral dip in Figure~\ref{boxes} remains relatively constant across the disk, though there is a decrease in depth along the SE direction.
This can be seen qualitatively in the images, as the surface brightness of the 3.08 $\mu$m Ice band in Figure~\ref{disks} is dimmer along the major axis (NE-SW).  The radial profiles along the NE-SW direction also show dimmer surface brightness in the 3.08 $\mu$m ice line from $50-150$ AU compared to the opposite direction.  The SE direction shows the shallowest absorption feature.  Since the depth of the absorption feature depends on the scattering angle \citep{Tazaki2021} for anisotropic scattering, this could be a result of the inclination of the disk, since the near-side of the disk is the SE.

The scattered light absorption feature at 3.08 $\mu$m is an effect of the wavelength dependence of the albedo for optically thick disks \citep[][see their Fig 2]{Inoue2008}.  Dust albedo is dependent on both grain size and composition \citep{Mulders2012}.  The overall disk scattered light surface brightness, in turn, depends on both the dust albedo and the geometry of the disk, especially the flaring index, since it is the amount of starlight scattered off the disk as a function of the radius \citep{Mulders2013}.  We find the absorption feature to be strongest in the   
outer regions along the major axis (NE-SW), where the flaring is the most pronounced, which could imply more icy grains at the disk surface.

\subsection{Spectro-Photometry} \label{sec4.2:spectra}
We construct three band scattered light photometric spectra along the major and minor axes in the NE (position angle PA$=54^\circ$), SW (PA$=234^\circ$), and SE (PA$=144^\circ$) directions.  The images are deprojected by the assumed inclination \citep[$i=23.2^\circ$;][]{Tang2017} in order to cleanly extract profiles without inclination biases.  The NW direction was excluded due to apparent negative flux (a result of disk oversubtraction) in the $L'$ band reduction in that direction.  

\begin{figure*}[tb]
\centering
\includegraphics[width=\linewidth]{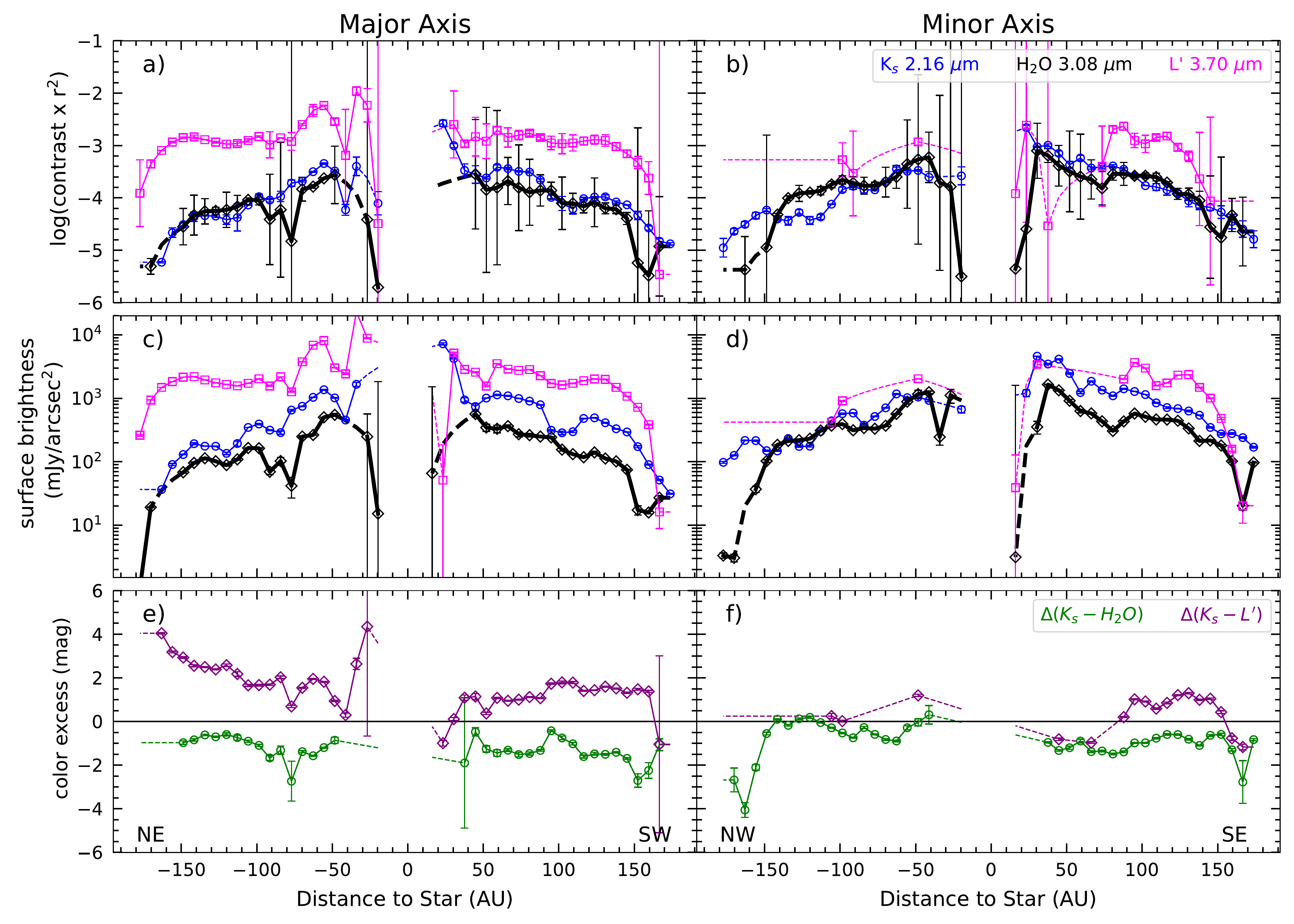}
 \caption{Radial profiles along the major (left) and minor (right) axis of the disk of AB Aur.  
 $K_s$ band is shown in blue, 3.08 $\mu$m Ice is shown in black, and $L'$ is shown in pink. \textbf{top)} Disk to star contrast of the observations as a function of distance from the central star. \textbf{middle)} Surface brightness of the disk as a function of distance from the central star. \textbf{bottom)} Scattered light color excess as a function of distance from of the central star.  $\Delta(K_s - \mathrm{H_2O})$ is shown in green, while $\Delta(K_s - L')$ is shown in purple.}
 \label{profiles}
\end{figure*}

\begin{figure*}[tb]
\centering
\includegraphics[width=\linewidth]{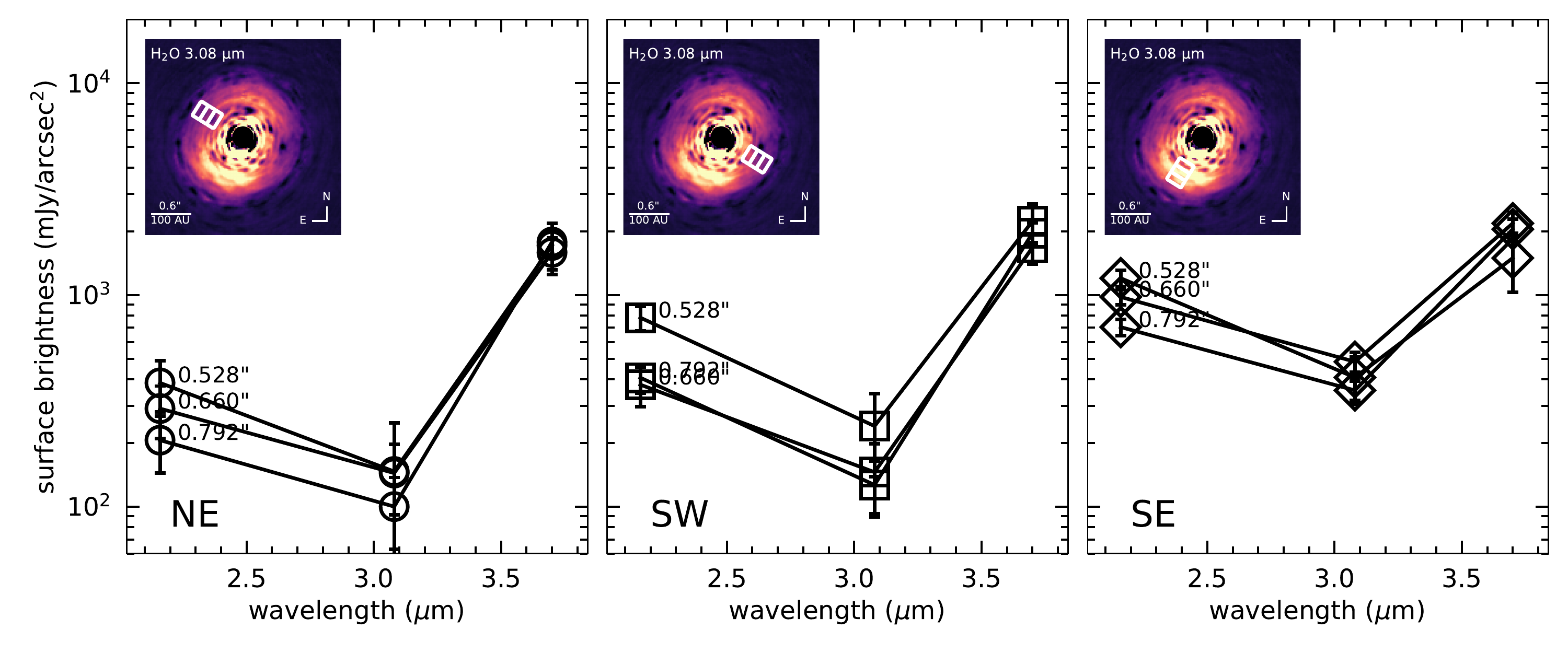}
 \caption{Surface brightness spectra along with NE (right), SW (center), and SE (left) direction of the disk of AB Aur.  Each spectra is extracted within a 0\farcs132$\times$0\farcs0.066 wide rectangle.  The spectra are centered at 0\farcs528, 0\farcs660 and 0\farcs792 along each axis. The inset 3.08 $\mu$m Ice band deprojected images shows the location of each box along each direction.  In all cases, a dip at 3.08 $\mu$m is seen.}
 \label{boxes}
\end{figure*}

As shown in Figure \ref{boxes}, spectra are extracted at 4 radial separations along each axis in 0\farcs132$\times$0\farcs0.066 wide rectangles. These rectangles are centered at 0\farcs528, 0\farcs660 and 0\farcs792.  The inset images in each panel of Figure~\ref{boxes} show the location of each extracted region overlain on the 3.08 $\mu$m disk image.  We see a dip in flux at 3.08 $\mu$m relative to $K_s$ and $L'$ at all distances from AB Aur. This deficit is not present in the spectrum of the central region (Figure~\ref{SED}), suggesting that the presence of icy grains leading to decreased scattering efficiency at this wavelength.
As evidenced by Figure~\ref{profiles}e and f, the slope from $K_s$ to $L'$ is consistently positive in the NE, SW, and in the SE and grows steeper at larger radial distances in the NE.  
In the southern portion of the disk the $\Delta(K_s - L')$ color averages around $1.08\pm0.40$ mag from 50-150 AU, while in the NE it increases from an average color of $1.61\pm0.44$ mag to a maximum color of $4.04\pm0.02$ mag by 150 AU away from the central star.  This strong red color is seen in the steep slope by 0\farcs792 in the NE spectra in Figure~\ref{boxes}.  Compared to the stellar $K-L$ color of $0.99\pm0.13$ mag, the disk is consistently redder indicating we are seeing more than just stellar radiation influencing the disk color.

\begin{figure}[tb]
\centering
\includegraphics[width=\linewidth]{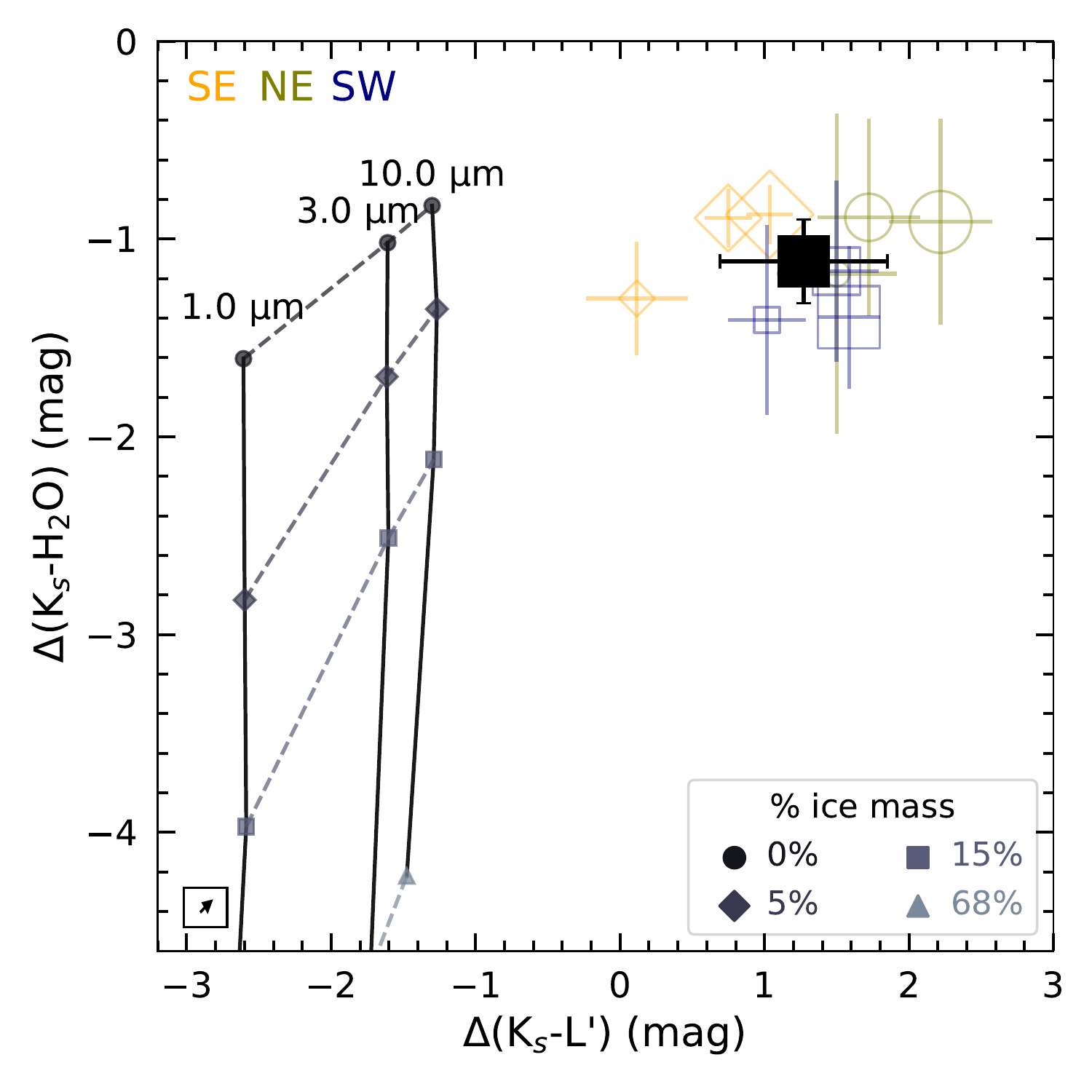}
 \caption{Color-color difference diagram of the MCFOST scattered light spectra from Figure \ref{boxes}.  The black lines are models for maximum grain sizes of 1.0, 3.0, and 10.0 $\mu$m (left to right), while the dashed lines show the ice mass fraction (circle - 0\%, diamond - 5\%, square - 15\%, triangle 68\%.)    
 The individual colored markers show the color difference for each extracted spectra (SE: orange diamond, NE: olive circle, SW: blue square) from Figure \ref{boxes}, with the size increasing further with distance from the star.  The black square indicate the average of all spectra.  The black arrow is the interstellar reddening vector.      }
 \label{colorcolor}
\end{figure}

Following \citet{Honda2009}, \citet{Honda2016} and \citet{Tazaki2021}, we determine the extent to which the observed 3.08 $\mu$m absorption is consistent with the presence of water ice grains, as well as the physical properties of these icy grains, using a color-difference diagram \citep{Inoue2008}.
The original model predictions derived in \citet{Inoue2008} are calculated for the $K$, $L'$, and H$_2$O ice ($3.02-3.16$ $\mu$m rectangular) filters on Subaru/CIAO instead of LBT/LMIRCam, and as they assume isotropic scattering \citep{Tazaki2021}, may overpredict ice abundance as well.  We therefore compare our observations to MCFOST anisotropic scattered light disk model spectra.  The method to compute MCFOST models is described in Section~\ref{Ice_and_grain}.
To derive $K_s$, H$_2$O ice, and $L'$ magnitudes from the MCFOST models (in order to calculate color), we convolve the model spectra with LBTI/LMIRCam filter profiles (as discussed in Section~\ref{photocal}) for each filter.
As we only want to see differential scattering imposed by the grains themselves, and not the intrinsic colors of the source of scattering photons (AB Aur), we remove its contribution following \citet{Honda2009} as:

\begin{align}
    \Delta (K_s - \mathrm{H_2O}) &= (K_s - \mathrm{H_2O})_{sca} - 
    (K_s - \mathrm{H_2O})_{source} \\
    \Delta (K_s - L') &= (K_s - L')_{sca} - (K_s - L')_{source},
\end{align}
where $sca$ represents the color of scattered light and $source$ is the intrinsic color of AB Aur.  
We show model predictions, observed colors for each region, as well as the combined average of all regions in Figure~\ref{colorcolor}.  We have corrected for extinction \citep[A$_V$ = 0.431; ][]{Vioque2018} using the extinction law of \citet{Cardelli1989}.   

While the observed $\Delta (K_s - L')$ colors at all computed locations in the disk lie redward of the predicted models, they are closest to models of an optically thick disk with a $0-15$\% ice mass fraction for grain sizes $>3\ \mu$m.   
This prediction is in agreement with grains sizes predicted by \citet{Perrin2009} based on their detailed scattered light modeling and within the large range ($0.1-40\ \mu$m) predicted by \citet{Bouwman2000} for a cold $50-400$ K disk from $28-175$ AU.   
Most of the grains are bluer in $\Delta (K_s - L')$ in the SE direction, and redder in the NE.  We find that the colors along the major and minor axes vary in $\Delta (K_s - L')$ color; the NE direction is redder compared to the southern part of the disk.  

The original models by \citet{Inoue2008} did not utilize a grain size distribution, assuming instead a ``typical" particle size. 
\citet{Tazaki2021} found that the grain-size distribution plays an important role in explaining both the depth of absorption and red $K_s-L'$ scattered light colors of previous observations.  When they rederived the \citet{Inoue2008} results using anisotropic scattering, they found that a) a log-normal size distribution can better explain redder scattered light, and b) the wavelength dependence of the phase function in turn helps increase the depth of the ice absorption feature for larger ($>1\ \mu$m) grains.  This is due to the decreased amount of the angular distribution of the light intensity at 3.08 $\mu$m compared $K$ and $L'$ for scattering angles $>30^\circ$.  However, though the models utilize Mie scattering, our results still lie redward of the model extremes in $\Delta(K_s - L')$.

\subsection{Ice Abundance and Grain Size} \label{Ice_and_grain}
We compare the observed scattered light photometric spectra to model spectra for various size distributions, grain sizes, and ice abundances.  We use the fixed model parameters from Table~\ref{tab3:fixedMCFOST}), and an inclination of 23.2$^\circ$ \citep{Tang2017}, PA of 328$^\circ$ \citep{Tang2017}, and scale height of 14 AU \citep{Perrin2009}.  We consider three different maximum grain sizes: 1.0, 3.0, and 10.0 $\mu$m and four different grain size distribution power law indices ($q$; $dN/da \propto a^q$): $-2.0$, $-2.5$, $-3.0$, $-3.5$. Changing the grain size distribution, should, in effect, change the amount of large grains and therefore the spectral slope.  
We use the same astronomical silicates and ice refractive indices, mass abundances, and material densities as in \citet{Inoue2008}.  The refractive indices for the astronomical silicates are from \citet{Draine1985} while the ices are from \citet{Irvine1968}.  The components are mixed using an effective mixing theory following the Bruggeman mixing rule to obtain the effective optical index of the new ``mixed" grain.  The optical properties are then obtained via Mie theory.  We assume a porosity of 60\% \citep{Perrin2009} and an inner radius of 0.3 AU .  To change the relative mass abundance of ices to silicates, we adjust the mass fraction within each grain so that there are either 0\%, 5\%, 15\% or 68\% water ice by mass, and the rest composed of astrosilicates.  
In order to confirm the validity and global energy budget of the models, we validate the modeled SEDs against the observed SED of AB Aur (see Figure~\ref{SED}).  We do not resolve the innermost disk emission (within 0.3 AU, $<<$ 1 pixel) in our LMIRCam data.  Therefore, we have added a 1900 K blackbody inner disk component \citep{Tannirkulam2008} to the stellar spectrum to include this additional excess NIR flux seen in the observed spectrum.     

\begin{figure*}[!tb]
\centering
\includegraphics[width=.86\linewidth]{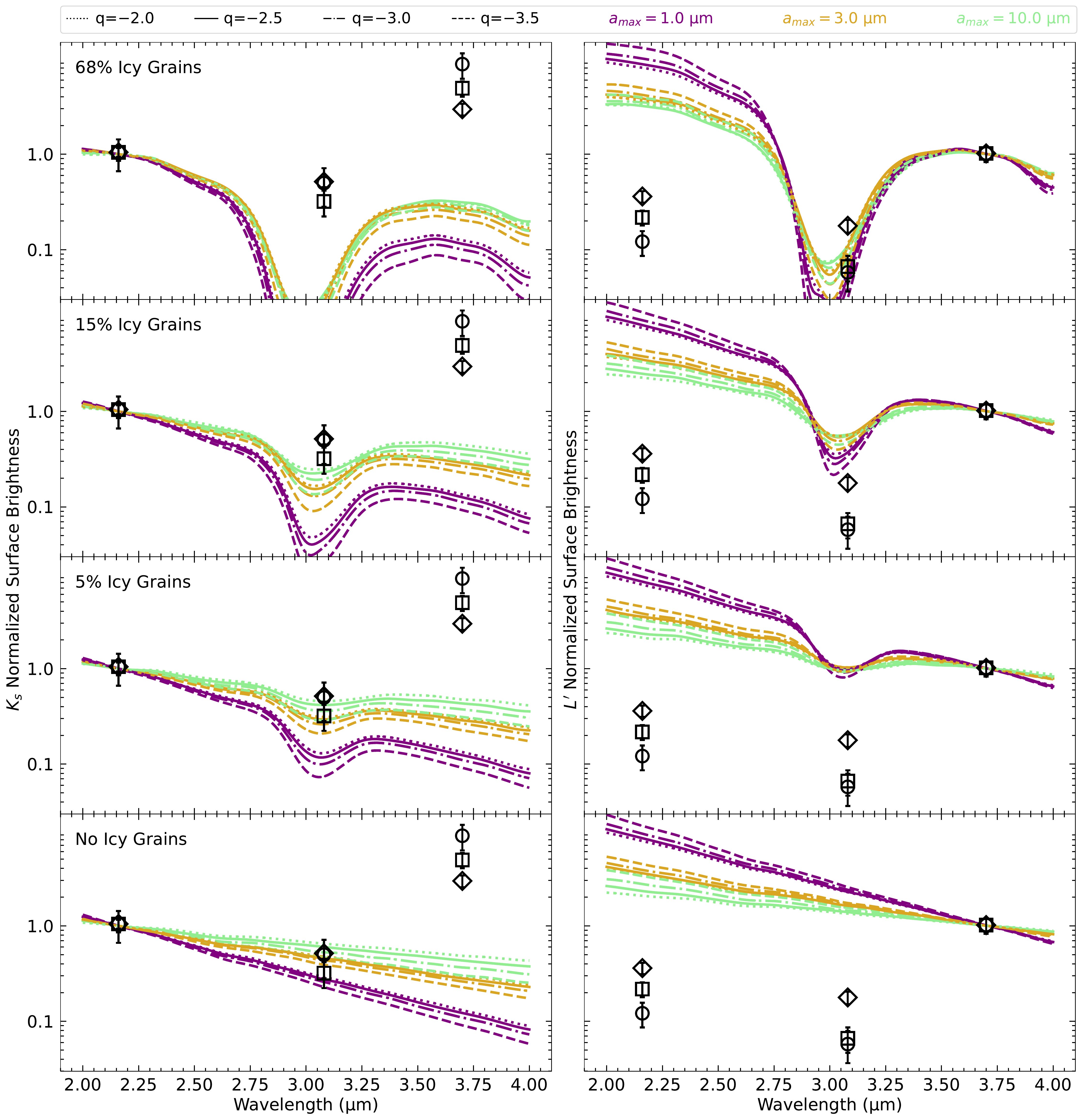}
 \caption{Scattered light spectra at a radius of 0\farcs792 normalized to $K_s$ band (lefthand column) or $L'$ band (righthand column) overlain with modeled spectra for various ice abundances.  The points indicate the relative photometry in the SE (diamond), NE (circle), and SW (square) directions.  Models with maximum grain sizes of 10 (green), 3.0 (gold), and 1.0 $\mu$m (purple), are shown overplotted for a range of power law grain size distribution indices, namely: $-$2.0 (dotted), $-$2.5 (solid), $-$3.0 (dash-dotted), $-$3.5 (dashed). Vertically, the panels range from high to low ice mass composition fractions, with values of 68\%, 15\%, 5\%, and 0\% ice. Grains that are 0\% or 5\% ice best match the $K_s$-$L'$ slope (lower left panels), while the 3.08 $\mu$m-$L'$ slope is best matched by grains with very high ice fractions ($\sim$68\%, upper right panel). } 
 \label{grainsize_iceamount}
\end{figure*}

We synthesize scattered light images of the disk using MCFOST for this small grid ($q$, $a_{max}$, and ice mass fraction) from $2-4\ \mu$m with a resolution of 0.1 $\mu$m.  Each image is convolved with a Gaussian PSF with FWHM equal to the FWHM of the nearest LMIRCam data (since the observed reference PSF stars are only in 3 distinct wavelength bands).  We then extract the scattered light spectra at 0\farcs792 from these model images.  Finally, we interpolated over these spectra in order to synthesize a continuous spectral model; the results are shown in Figure~\ref{grainsize_iceamount}.  

For grains composed of $0 - 68$\% water ice by mass, the mass abundance does not greatly affect the $K_s-L'$ color; however, the deficit at 3.08 $\mu$m increases by almost an order of magnitude as the icy grain abundance increases from 0 to 15\% icy grains.  We also find that an increase in maximum grain size has the opposite effect, causing the absorption at 3.08 $\mu$m to decrease. 
Decreasing the size distribution power law index does redden the spectral slope; however, its effect is small relative to the change in maximum grain size, which reddens the slope by almost 50\% at $L'$ from 1 $\mu$m to 10 $\mu$m.

\begin{figure}[htp]
\centering
\includegraphics[width=\linewidth]{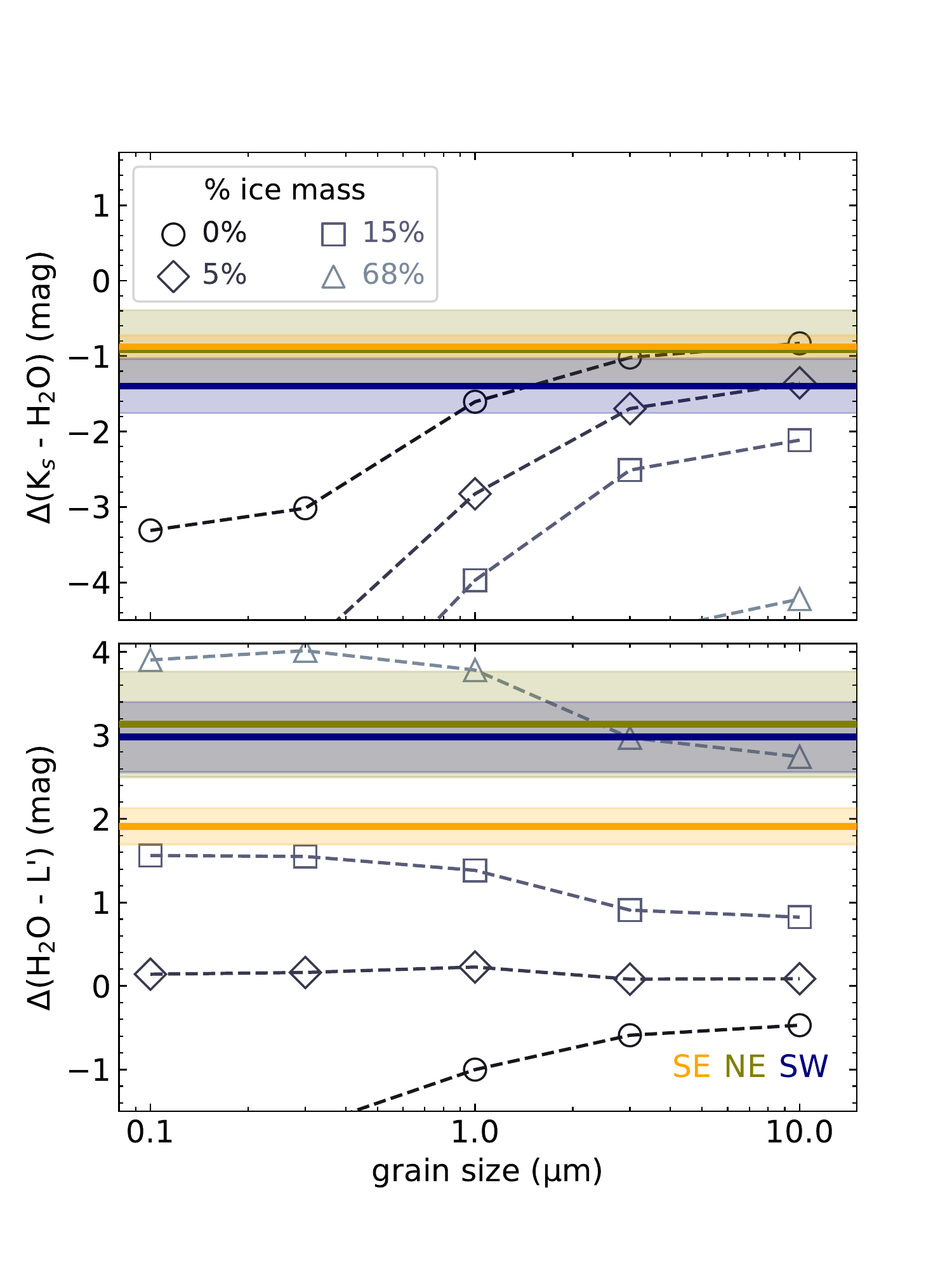}
\caption{$\Delta (K_s - \mathrm{H_2O})$ color (top) and $\Delta (\mathrm{H_2O}-L')$ color (bottom) of scattered light spectra as a function of grain size.  The dashed lines show the ice mass fraction for MCFOST models with grain size distribution power law index $q=-3.5$.  The solid horizontal lines are the colors of the AB Aur disk with LMIRCam (SE: orange, NE: olive, SW: blue) in Figure~\ref{grainsize_iceamount}.  The best-fitting ice mass fraction differs substantially between the two colors from $0-15$\% in $\Delta (K_s - \mathrm{H_2O})$ to $\sim68\%$ in $\Delta (\mathrm{H_2O}-L')$.    } 
\label{colorgrainsize}
\end{figure}

From Figure \ref{boxes}, we find that the $\Delta(K_s - \mathrm{H_2O})$ slopes are relatively consistent.  Therefore, to compare the observed spectra to the above models, we show the spectra at 0\farcs792 for each direction.  We first normalize the spectra to $K_s$ (left panel of Figure~\ref{grainsize_iceamount}) and find that 
the observed surface brightness from $K_s$ to $\mathrm{H_2O}$ in the SW (square) direction can be reproduced with a model with $5$\% icy grains by mass with a grain size $>3\ \mu$m and $q=[-2.0, -3.5]$ ($r^2 = 0.98$). 
In the SE (diamond) and NE (circle) direction, a model with 5\% icy grains by mass with a grain size $>$ 10.0 $\mu$m and $q=[-2.0, -3.0]$ can reproduce the observed surface brightness ($r^2\sim 0.91$).  However, models with no icy grains can also reproduce observed ice surface brightness.  This is easily seen in the top panel of Figure~\ref{colorgrainsize}, where the $\Delta (K_s - \mathrm{H_2O})$ color is most consistent with models with grains composed of $0-5\%$ ice by mass and grain sizes $\gtrsim3\ \mu$m. We discuss the implications of having a grain population composed of mostly large grains below and in Section~\ref{red light spectra}.

These results could indicate that the observed disk surface is ice-poor and primarily composed of larger grains, perhaps due to intense photoevaporation or turbulence.  
None of the models we constructed were able to reproduce the observed red $\Delta(K_s - L')$ color while normalized to $K_s$ band, indicating the power law size distribution may not be sufficient to describe this slope.  The same conclusion was made by \citet{Tazaki2021}, who found that a log-normal distribution reproduces red scattered light profiles. This is because the power law size distribution is dominated by 1 $\mu$m grains, causing the scattered light spectra to become almost gray \citep[the albedo and scattering cross section are wavelength independent for $\gtrsim1\ \mu$m grains][]{Tazaki2021}.  If we instead assume a log-normal distribution will match the observed slope, then \citet{Tazaki2021} claims that a top-heavy power law (e.g. smaller power law index) distribution should give the closest match to the log-normal distribution as there will be more large grains.  However, as shown in Figure \ref{grainsize_iceamount}, even decreasing the power law index does not redden the slope enough, indicating the redward slope might have another cause (see Section~\ref{sec:opticaldepths}).

Therefore, we also normalize the spectra to $L'$, as shown in the right panel of Figure~\ref{grainsize_iceamount}, to quantify the fraction of icy grains necessary to match the $\Delta(\mathrm{H_2O}-L')$ slope.  We find that only grains that are $\gg5$\% water ice by mass can reproduce the observed spectral slope.  This is highlighted in the bottom panel of Figure~\ref{colorgrainsize}, where the color is consistent with models with grains $\sim 68$\% ice by mass and grain sizes $>1\ \mu$m .  As we discuss in Section \ref{sec:discussion}, this composition is unlikely within 200 AU, lending credence to our assertion that the $L'$ scattering surface is likely at a different location in the disk surface layer.

\begin{figure}[htp]
\centering
\includegraphics[width=\linewidth]{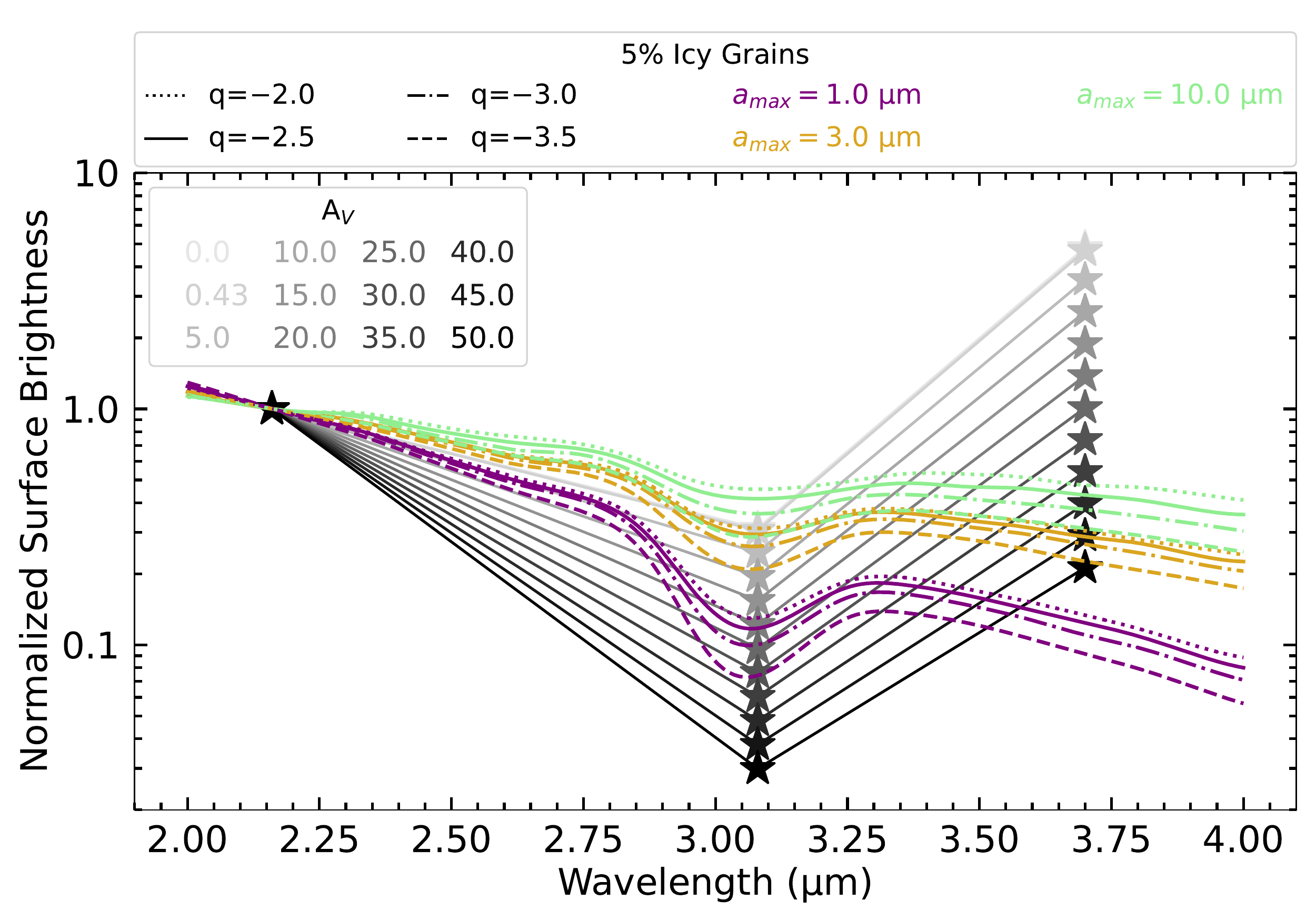}
\caption{Effect of dereddening on the AB Aur spectra in order to match the modeled scattered light spectra.  The darker colors indicate a larger dereddening correction ($A_V$).  The models are as in Figure~\ref{grainsize_iceamount} for an ice mass composition fraction of 5\% ice. $L'$ must have an extra $\sim$45-50 mags of extinction to get its observed spectral slope.}
\label{extinction}
\end{figure}

\section{Discussion} \label{sec:discussion}
We have found a shallow observed dip at the 3.08 $\mu$m ice line relative to $K_s$, and a deeper observed dip relative to $L'$, suggestive, though unlikely, of more icy grains by mass.  When compared to models, the shallow dip relative to $K_s$ suggests that the surface is very ice-poor with a ice mass fraction of $\sim0-5$\% if the grains are at least 3 $\mu$m in size.

\subsection{Comparison to HD 142527 and HD 100546}
As only two other disks, HD 142527 and HD 100546, have been studied with the water ice deficit technique \citep{Honda2009, Honda2016, Tazaki2021}, we discuss the similarities among the three disks below, with their relevant properties listed in Table~\ref{comparetoHDdisks}.  Like AB Aur, HD 142527 and HD 100546 are both Herbig Ae/Be stars with strong absorption at 3.09 $\mu$m, as well as a red $\Delta(K - L')$ slope.

HD 100546 shows changes in inferred feature depth as a function of distance from the star.  \citet{Honda2016} found that the optical depth increases from $\sim0.5$ to $\sim1.0$ further from the star, potentially due to an increase in ice grains outward in the disk.  They also find strong asymmetry in the ice absorption optical depth along the minor axis of the disk, which they attribute to the disk geometry; larger optical depth for the near side of the disk.  However, \citet{Tazaki2021} found the opposite to be true, concluding that this asymmetry must be due to ice abundance variation.  AB Aur does not have a strong asymmetry; there is an increased optical depth along the major axis which decreases along the minor axis.  The optical depth also decreases closer to the central star, which could be due to a decrease in icy grains.  These disks show remarkably similar ice abundances and ice line optical depths that could merely be due to their similar evolution and stellar properties.  

\begin{deluxetable}{lcccc}[htp]
\centering
\tabletypesize{\scriptsize}
\tablecaption{Comparison of Disk Properties
\label{comparetoHDdisks}
}
\tablehead{\colhead{} & \colhead{AB Aur} & \colhead{HD 142527} & \colhead{HD 100546} & \colhead{Refs\tablenotemark{$a$}}
}
\startdata
inclination (deg) & 23.2 & 28 $\pm$0.5 & 32-42 & (1)/(2)/(3)\\
age (Myr) & 4$\pm$1 & $\sim$1 & $>10$ & (4)/(5)/(6) \\
avg $\Delta(K-\mathrm{H_2O})$ & -1.11$\pm$0.22 & -1.01$\pm$0.41 & -- & (7)/(8)/-- \\
avg $\Delta(K-L')$ & 1.27$\pm$0.61 & 0.44$\pm$0.28 & -- & (7)/(8)/-- \\
ice/silicate & 0-0.05 & 0.06-0.2 & 0.06-0.2 & (7)/(9)/(9) \\[-1ex]
\hspace{0.1cm} mass ratio & & & & \\
\enddata
\tablenotemark{a}{Reference for AB Aur/HD 142527/HD 100546}
\tablerefs{(1) \citet{Tang2017}; (2) \citet{Perez2015}; (3) \citet{Pineda2019}; (4) \citet{Rodriguez2014}; (5) \citet{Fujiwara2006}; (6) \citet{Grady2001}; (7) this work; (8) \citet{Honda2009}; (9) \citet{Tazaki2021} } 
\end{deluxetable} 

\citet{Honda2009} found a ice to silicate mass ratio of $>$ 2.2 along the SW of the disk of HD 142527, though \citet{Tazaki2021} attributes this high ratio to the assumption of isotropic scattering in the \citet{Inoue2008} models used to interpret the data.  With the inclusion of anisotropic scattering, \citet{Tazaki2021} find a mass ratio of $0.06 - 0.2$, similar to the ratio of 0.05 found in AB Aur.  They also find a reddish scattered spectral similar to AB Aur, through the disk of AB Aur is far redder at $L'$ band compared to HD 142527 (see Table~\ref{comparetoHDdisks}).   
The predicted surface snowline of HD 142527 is located between $100-300$ AU according to models of photodesorption from far-ultravoilet photons from \citet{Oka2012}.  \citet{Tazaki2021} attributes the low ice abundance to being near the surface water snowline. Overall, these disks are quite similar in ice abundance, composition, and color.  More work needs to be done to understand if these are typical or unique in their grain properties among Herbig Ae/Be stars. 

\subsection{Red Scattered Light Spectra}\label{red light spectra}

Though the red $\Delta(K_s - L')$ scattered light color that we see in AB Aur has also been seen in the disks of HD 142527 and HD 100546, blue NIR to $L'$ disk colors have also been detected \citep[e.g. GG Tau;][]{Duchene2004}. This extreme color variation among disks at similar evolutionary stages highlights the complex nature of scattering and disk surface brightness.  Differences in grain properties (grain size, porosity, vertical dust settling, polarization fraction), as well as wavelength dependencies of the albedo and scattering phase function can all affect the NIR colors of the disk \citep{Duchene2004, Mulders2013, Tazaki2019}.  Therefore, we review a range of factors that could explain the red scattered light color of AB Aur below.

We first quantify how much local extinction, beyond the 0.431~mag observed interstellar extinction for AB Aur \citep{Vioque2018}, would be necessary to produce the observed $\Delta(K_s - L')$ slope.
As shown in Figure~\ref{extinction}, if we deredden the observed spectra by various values of $A_V$, we are able to match the modeled extinction free slope.  However, we find that this requires $A_V \sim 40-50$ mags, which is nonphysical. Therefore, other explanations beyond local extinction are required to reproduce the red $\Delta(K_s - L')$ scattered light color.

\citet{Tazaki2021} attributes the red scattered light color of AB Aur to the spiral structure in the disk \citep{Hunziker2021}.  One possibility for the formation of these spirals arms is that they result from gravitational instabilities (GI) in the disk. GI can help bring large grains to the surface of the disk, which then forward scatter light at short wavelength thereby reducing the scattering efficiency of grains and causing the red slope \citep{Tazaki2019, Tazaki2021, Mulders2013}. 
Recent studies \citep{Dullemond2019, Kuffmeier2020} claim that the spiral arms around AB Aur are formed from late infall of material on the disk, and therefore, there should be blue-scattering only in the NIR \citep{Tazaki2021}. 

However, there are two caveats that can help explain our redward slope.  First, previous observations of the disk of AB Aur have not gone as red as $L'$ band.  
Therefore, while the disk of AB Aur shows a blueward slope in the bluer end of the NIR supporting the claim of blue-scattering, the spiral arms in $L'$ are quite prominent and could help influence the slope at the redder end of the NIR.   
Second, the most prominent spiral arms in AB Aur extend outward from 200 AU to 400 AU \citep{Fukagawa2004, Boccaletti2020} in the outer disk.  The spirals we see in $L'$ are within 200 AU, in the innermost part of the disk and could therefore not be part of the later infalling material, and instead be formed from gravitational instability \citep{Cadman2021}.  Therefore, the reddish scattered light in the inner disk of AB Aur could be explained by the spiral arms effectively diffusing large grains to the disk surface, similar to HD 142527 and HD 100546. 

\citep{Hunziker2021} found that for HD 142527, the red scattered light spectrum of the disk could be explained by a simple scattering model with strong forward scattering induced by large, highly porous aggregate dust grains.  They also found that this dust model was able to explain the high polarization fraction seen in HD 142527.    Due to to the strong similarities between HD 142527 and AB Aur (disk inclination, age, spectral type), we postulate that a population of large porous aggregate grains could also help to explain the red $\Delta(K_s-L')$ color and the high polarization fraction \citep[$\sim40\%$;][]{Perrin2009} seen in the disk of AB Aur.
On the other hand, \citet{Tazaki2021b} found from scattered light disk simulations that while porous grains will have higher polarization fractions \citep{Kirchschlager2014, Min2012}, only large compact dust grains (with lower polarization fractions) can produce red scattered light. Detailed dust grain modeling of AB Aur is needed to disentangle the relationship between porosity, disk color, and polarization fraction.  

\subsection{Optical Depth Effects} \label{sec:opticaldepths}

We investigate the varying ice abundances necessary to explain the $\Delta(K_s-\mathrm{H_2O})$ and $\Delta(\mathrm{H_2O}-L')$ colors.  
The differences in ice mass fraction between $\Delta(K_s-\mathrm{H_2O})$ and $\Delta(\mathrm{H_2O}-L')$ colors along with the strong morphological differences could be a result of these wavelengths probing different depths of the disk atmosphere. This is supported by the fact that the $K_s$ and $L'$ images show marked differences in shape and morphology (notably the presence of spiral arm features, though their apparent absence in $K_s$ could be a result of its reduced brightness contrast to the surrounding disk at shorter wavelength along with its more limited sensitivity).  
If the opacity of the disk is reduced at $L'$, then scattered light at this wavelength will probe deeper into the disk, allowing us to image the spirals.  If this is the case, then the dust populations probed at $K_s$ and $L'$ are likely different in size and/or composition, as these properties should have radial and vertical gradients within the disk. This is our favored explanation for the ensemble of observations, as it explains both the apparent morphological differences and the fact that the three wavelength photometry is not well-matched to any theoretical predictions for icy grains.

Visually, we find that the $K_s$ and the 3.08 $\mu$m ice images are morphologically similar (Figure~\ref{contours}b), indicating they likely probe nearby components of the disk surface, though the ice component is most likely at a somewhat lower depth given its longer wavelength. The $L'$ morphology is markedly different from both $K_s$ and 3.08 $\mu$m, suggesting that it may be scattered from significantly deeper in the disk.   The discrepancy in the icy grain fractions required to match the $\Delta(K_s-\mathrm{H_2O})$ ($0-5$\%) and $\Delta(\mathrm{H_2O}-L')$ (68\%) colors lend additional support to this conclusion. Given that an ice to silicate mass ratio of 68\% is likely nonphysical on the disk surface (especially if there is strong photodesorption; see Section~\ref{sec5.4}), our hypothesis that the $K_s$ and 3.08 $\mu$m scattering surfaces are roughly co-located is further supported.  However, care must be taken when interpreting both the $K_s-L'$ and $\mathrm{H_2O}-L'$ slopes, and more work is needed to disentangle optical depth effects from $K_s$ to $L'$. 

\subsection{Temperature Dependence on the Survival of Surface Icy Grains} \label{sec5.4}
\citet{Riviere-Marichalar2020} derived the surface temperature of the disk of AB Aur using $^{12}$CO(1-0) gas kinematics.  They estimates a temperature of 70 K near the central star, decreasing to 10 K by 650 AU.  The average dust temperature across the disk is around $25-30$ K \citep{Pietu2005, Woitke2019}; though warmer than other similar disks (e.g. HD 142527), it is still below the temperature necessary for water ice desorption ($\sim 120-180$~K).  Due to this, the detection of ice absorption could be possible.  For these large icy grains to exist on the disk surface, turbulent diffusion or strong disk winds (via photoevaporation) must be evoked to explain their presence, as evidenced by \citet{Tazaki2021}.  

For Herbig Ae/Be stars, \citet{Agundez2018} found that the photodissociation rate of water ice at 100 AU is $1\times10^{-4}-1\times10^{-5}$ s$^{-1}$ ($\sim310-3100$ yr$^{-1}$).  This is driven by stellar radiation entering the disk at a low grazing angle in flared disks, causing it to be the dominant source of radiation on their surfaces.  By 4 Myr (approximate age of AB Aur), water at the surface of the disk is highly dissociated, and the vertical extent of icy grains is controlled by photodesorption, shifting the snowline down and inward (or pushing the surface snow line out).  Models by \citet{Oka2012} found photodesorption to be quite efficient for high temperature ($\mathrm{T}\sim10000$~K) and luminosity (L~$\sim40$~L$_\odot$) Herbig Ae/Be stars (see their Figure~9).  

At these high temperatures, they conclude that the surface snow line should be significantly further away from the central star, $> 1000$ AU, due to the high rate of photodesorption.  Icy grains should, in effect, not exist on the surface of disks around hot stars (where T$_* >$ T$_c$, the critical temperature between sublimation/condensation that controls the radius of the snowline and photodesorption/condensation mechanisms), as the ice-condensation front (where condensation of icy grains is equal to the rate of photodesorption) forms at the surface of the disk.  
AB Aur, with a temperature $\sim 9700$ K and luminosity of $\sim 40$ L$_\odot$, has a critical temperature of T$_c \sim 4500$ K assuming an average dust temperature of 36 K \citep{Woitke2019}, and should therefore have exceedingly strong photodesorption. As shown by Figure~\ref{grainsize_iceamount}, the $\Delta{(K_s - \mathrm{H_2O})}$ color aligns with models with $0-5$\% icy grains by mass (and is the more likely ice abundance) indicating surface water ice could be affected by photodesorption.  If we assume that the $K_s$ and 3.08 $\mu$m ice components are at approximately the same location on the disk surface, then the 0\% icy grains by mass within 150 AU is consistent with \citet{Oka2012}. However, though small, models with 5\% icy grains by mass do match the observed $\Delta{(K_s - \mathrm{H_2O})}$ color, indicating that if the ice surface is at a slightly lower disk surface depth compared to $K_s$, the icy grains at $\sim100$ AU could both exist and be shielded from some photodesorption on the surface.

\subsection{H$_2$O Snowline in Relation to Planet Formation in AB Aur}
Within the disk of AB Aur, two indirectly inferred planet candidates have been reported at 30 and 140 AU \citep{Boccaletti2020}.  The inner planet candidate mass is tentatively estimated at $4-13$ M$_{Jup}$ \citep{Boccaletti2020, Pietu2005, Tang2017}, while the outer is tentatively estimated at 3 M$_{Jup}$ \citep{Boccaletti2020}; the interpretation is indirect, based on a substructure that can be interpreted as a planet-induced perturbation (but other explanations are plausible; Jorquera et al. 2022, in press).  The estimated masses and locations of the planets are consistent with formation through gravitational instability \citet{Cadman2021}.  

Our detection of reduced ice absorption at $\sim100$ AU would place the potential outer companion within the region where we see little to no ice absorption.      
Far-infrared spectroscopic observations of AB Aur \citep{Fedele2013} did not detect H$_2$O emission, indicating that water molecules in the outer disk beyond the snow line have likely been photodissociated by stellar radiation.  \citet{Fedele2013} conclude that this would result in a lack of water reservoir in the outer disk necessary for giant icy planet core formation.  Since the location of the midplane snowline has not been measured, it is not certain if there are icy volatiles at this separation in the disk midplane.  Therefore, whether or not there are enough icy volatiles for icy core formation for an outer planet is unknown, and would require further observations and detections of volatile snow lines. 

\section{Conclusion} \label{sec:conclusion}
In this paper, we observe the (pre)transitional disk around AB Aur in order to explore the presence of water ice in the surface layers of the scattered light disk.  Using LBTI LMIRCam at $K_s$, the 3.08 $\mu$m water ice line, and $L'$, we look for a deficient of NIR scattered light at the ice line, as icy grains absorb preferentially at this wavelength.  We use the $\Delta(K_s - \mathrm{H_2O})$ and $\Delta(K_s- L')$ color differences in order to probe the grain size distribution and ice abundance within this disk. We find the following.
\begin{enumerate}
    \itemsep0em
    \item We detect the disk of AB Aur at high SNR in all three LBTI LMIRCam bands, $K_s$, 3.08 $\mu$m ice, and $L'$.  After removing the starlight using KLIP-RDI, the disk is visible out to 175 AU (1\farcs1).  We find strong morphological differences between $K_s$ and $L'$ band along with prominent $L'$ inner spiral arms.  The $K_s$ image closely resembles the morphology found in $H$ polarized intensity imaging.  The 3.08 $\mu$m ice band image has a closer morphology to $K_s$ than $L'$.
    \item After extracting scattered light photometry along the major and minor axes for each band, we detect absorption at 3.08 $\mu$m from $90-120$ AU.  Using a color-color difference diagram \citep{Inoue2008}, we argue that our observations are consistent with the presence of icy grains on the surface of the disk of AB Aur with sizes $3\ \mu$m.  
    \item By comparing to MCFOST models of varying ice abundances and maximum grain sizes, we find the $\Delta(K_s - \mathrm{H_2O})$ slope most closely matches a model with a ice/silicate mass ratio $\sim$ 0.05 (5\% icy grains by mass) with a maximum grain size of $3-10$ $\mu$m.  The reddish $K_s$ and $L'$ continuum slope can be explained by the presence of these large grains (though the MCFOST models cannot predict this slope).  However, these large grains can only exist on the disk surface due to either disk turbulence or winds (via photoevaporation) preventing grain settling.
    \item We argue that the strong $\Delta(K_s - L')$ slope seen in the scattered light spectra is most likely to be a result of probing different scattering surfaces at different depths within the disk with the $L'$ disk component probed at a deeper disk surface.  Further investigation is necessary to update model predictions of disk colors under the assumption of differential optical depths between NIR wavelengths.
\end{enumerate}

These results suggest that the surface of the disk of AB Aur is highly ice-poor resulting from strong photodesorption of large grains.  At the higher stellar temperature and luminosity of AB Aur, the surface snowline lies beyond 100 AU as the stellar radiation field pushes the ice condensation front outward along the surface, potentially as far as 1000 AU (far beyond our detection limits). The small ($0-5\%$) presence of icy grains at the disk surface is suggestive of vertical mixing in the disk or that the grains are shielded from photodesorption (due to optical depth effects) in order to survive. Therefore, if we want to strongly detect the surface snow line, sensitive observations of the outer disk, as well as millimeter observations of the midplane snow line, are necessary.

\acknowledgments
We thank Esther Buenzli, Vanessa Bailey, Timothy Rodigas, and Jared Males for contributing to or obtaining the original imagery. We thank the anonymous referee for their careful review. We thank Kellen Lawson for his helpful suggestion to look into the differential evolution algorithm as a way to fit the forward modeled disks. SKB and KBF acknowledge support from NSF AST-2009816.  
S.J. acknowledges support from the National Agency for Research and Development (ANID), Scholarship Program, Doctorado Becas Nacionales/2020 - 21212356.
L.P. gratefully acknowledges support by the ANID BASAL project FB210003, and by ANID, -- Millennium Science Initiative Program -- NCN19\_171.
We acknowledge the use of the Large Binocular Telescope Interferometer (LBTI) and the support from the LBTI team. 
MCFOST is funded from the Australian Research Council under contracts FT170100040 and DP180104235, from Agence Nationale pour la Recherche (ANR) of France under contract ANR-16-CE31-0013.
This research has made use of the SVO Filter Profile Service (http://svo2.cab.inta-csic.es/theory/fps/) supported from the Spanish MINECO through grant AYA2017-84089.  
This publication makes use of data products from the Wide-field Infrared Survey Explorer, which is a joint project of the University of California, Los Angeles, and the Jet Propulsion Laboratory/California Institute of Technology, funded by the National Aeronautics and Space Administration.

\facilities{LBT(LMIRCam)}

\software{pyKLIP \citep{Wang2015},
          diskFM \citep{Mazoyer2020}, 
          MCFOST \citep{Pinte2006, Pinte2009}, 
          Astropy \citep{astropy:2013, astropy:2018}, numpy, scipy, matplotlib \citep{Hunter2007}, 
          Astrolib PySynphot \citep{pysynphot},
          Dewarp \citep{Spalding2019}, DebrisDiskFM \citep{Ren2019}, pymcfost (https://github.com/cpinte/pymcfost), mcfost-python (https://github.com/swolff9/mcfost-python)
          }

\bibliography{ref}
\bibliographystyle{aasjournal}

\newpage

\appendix 
\section{Geometric Disk Modeling} \label{appendix:diskmodel}
We use previously measured AB Aur observations and model parameters to fix the majority of the non-geometric parameters \citep[e.g. stellar temperature and luminosity, dust mass and composition, grain porosity and radius, grain size power law etc.;][]{Perrin2009, diFolco2009, Tang2017}.  Following \citet{Perrin2009} and \citet{diFolco2009}, we use a mixture of astrosilicate grains \citep{Draine1985, Weingartner2001} with 60\% porosity from $0.005-1\ \mu$m and with grain sizes described by the power law $dN/da \propto a^{q}$, where the power law index ($q$) is $-3.5$.  We set the inner radius to 0.2 AU \citep{Perrin2009}, and the outer radius to the outer edge of each band's KLIP-RDI image ($K_s=189$ AU, 3.08~$\mu$m$=168$ AU, $L'=175$ AU).  See Table~\ref{tab3:fixedMCFOST}.
As the purpose of modeling the disk of AB Aur at each wavelength is to determine the throughput and reproduce the observed surface brightness, our primary concern is finding a model that reproduces the brightness and morphology of the disk, and derived parameters may be unphysical. 

We investigate four different geometric parameters to achieve a best fit: inclination, ($i$), position angle of the disk along the minor axis (PA), scale height at 100 AU ($h_{100}$), and flaring index ($\beta$; h(r) = $h_{0}(r / r_0)^\beta$ where $r_0=100$ AU and $h_0 = h_{100}$).  Though all of these parameters have been previously fit for AB Aur, our initial tests (fitting the bands both simultaneously and separately) using best fit values from other work ($i = 23.2^\circ$ \citep[ALMA;][]{Tang2017}, PA = 328$^\circ$ \citep[ALMA;][]{Tang2017, Perrin2009}, $h_{100} = 14$ AU \citep{Perrin2009} or $h_{100} = 8$ AU \citep{diFolco2009}, $\beta=1.3$ \citep{Perrin2009, diFolco2009}) provided ill-fitting models. This led us to model the disk at $K_s$, 3.08 $\mu$m, and $L'$ separately, and without limiting ourselves to physically-motivated parameters.  

To forward model, we create 2.16 $\mu$m, 3.08 $\mu$m, and 3.70 $\mu$m scattered light models of an AB Aur analog disk using the MCFOST Monte Carlo radiative transfer code \citep{Pinte2006, Pinte2009}, which utilizes Mie theory. The MCFOST model is then convolved by the unsaturated PSF of AB Aur and multiplied by an additional fit parameter, a scaling parameter ($A$).  This scaled and convolved image is then forward modeled using the \texttt{pyKLIP} DiskFM package \citep{Mazoyer2020} to produce a simulated KLIP-RDI reduced disk model under the same KLIP parameters as the observations.

To find the best fit model, we tested a variety of optimization methods, including a Markov-Chain Monte-Carlo (MCMC) and a coarse grid search.  The near face-on nature of the disk, together with the distinct and complex substructure between the three bands, indicates parameter degeneracies. Local minima, as well as differences in geometric parameters between bands, is both possible and expected.  

We first attempted to fit the three wavelengths separately using MCMC by combining the \texttt{pyKLIP} diskFM implementation from \citet{Mazoyer2020} and \citet{Chen2020} with the MCMC implementation \citep{debrisdiskFM} from \citet{Ren2019} and \citet{Ren2021}.  However, due to the degeneracy between several variables and a broad parameter space, the MCMC walkers could not find global minima that were qualitatively good fits to the data.  

We next attempted a coarse grid search across the four geometric free parameters ($i$, PA, $\beta$, $h_{100}$) and the scaling amplitude ($A$).  The range in parameter space for each free parameter was determined either from previous literature estimates ($\beta$, $h_{100}$, PA) or from initial model testing ($i$, $A$).  We initially required the inclination to be fixed at 23.2$^\circ$, the inclination found from ALMA dust continuum kinematics \citep{Tang2017}; however, this inclination produced a significantly poor fit to the disk at all three wavelengths, while more face-on inclinations ($i<15^\circ$) produced better fits.  Though our coarse grid (5 parameters, 3$-$20 values for each parameter) found best-fit models, these models were dependent on the grid structure.  To expand to a finely sampled grid would be computationally time consuming and would expand to an unfeasible number of models. 

\begin{deluxetable}{lcc}[bt]
\centering
\tabletypesize{\scriptsize}
\tablecaption{MCFOST Fixed Parameters \label{tab3:fixedMCFOST} }
\tablehead{\colhead{Parameter} & \colhead{Value} & \colhead{References}
}
\startdata
$T_{\rm eff}$ (K) & 9772 & (1) \\
$R$ ($R_\odot$) &2.5 & (2)\\
$M$ ($M_\odot$) & 2.4 & (2)\\
$R_\mathrm{inner}$ (AU) & 0.2 & (1) \\
$R_\mathrm{outer}$ (AU) & 175 &  \\
$M_\mathrm{dust}$ ($M_\odot$) & $1 \times 10^{-4}$ & (1) \\
$a_\mathrm{max}\ (\mu$m) & 1.0 & (1)\\
$q$ & $-$3.5 & (1) \\
Dust Composition & Astrosilicates & (1, 3, 4)\\
\enddata
\tablerefs{(1) \citet[][and references therein]{Perrin2009}; (2) \citet{DeWarf2003}; (3) \citet{Draine1985}; (4) \citet{Weingartner2001} }
\end{deluxetable} 
Therefore, following \citet{Lawson2020}, we turned to a combination of these approaches via a differential evolution optimization \citep[DE;][]{Storn1997}.  For a detailed description of the DE algorithm and its application to modeling scattered light disks, see \citet{Lawson2020}.  In short, DE is a global optimization and search algorithm that can quickly and efficiently explore multi-dimensional, correlated, non-differential and/or nonlinear  parameter spaces to find a global minimum.  By only exploring within specified parameter boundaries, DE starts by initializing a population of models randomly sampling the parameter space. These models then iteratively evolve (or mutate) towards solutions that best fit the data.  This evolution allows the walkers to explore multiple local minima and converge on the global minimum relatively quickly with a smaller (compared to a grid search or MCMC) number of samples.  Following \citep{Lawson2020}, we initialize $N_{pop}=50$ models, 10$\times$ the number of free parameters (M=5).  These models are then forward modeled using \texttt{pyKLIP} diskFM, as discussed above, and the forwarded models are compared to the reduced observations by computing 

\begin{equation}
    \chi_r^2 = \frac{\chi^2}{\nu} = \frac{1}{\nu}\sum_S \frac{\rm (Data - Forward\ Model)^2}{\rm Uncertainty^2}
\end{equation}

following \citet{Chen2020}, where $\nu$ is the number of degrees of freedom given by N-M (N is the number of pixels within $S$), and $\chi^2$ is computed over $S$, a circular annulus.  The uncertainty is measured as discussed in Section~\ref{sec2.2:data reduction}.  Then, the population mutates following the "best/1/bin" strategy of \citet{Storn1997} with the mutation constant in the range [0.5, 1.0]. Each mutation replaces its "parent" if a combined version of the two models (determined by changing select individual parameters of the parent with the same parameters of the mutant if a random probability P$\geq$0.7 is found for that parameter) yields a better fit (e.g. smaller $\chi_r^2$).  This mutation, recombination, and replacement is continued until the population converges to a single solution with the global minimum $\chi_r^2$.  Since the population is initialized across the whole grid, the algorithm can robustly explore across local minima without getting trapped. 

\begin{figure}[tb]
    \centering
    \includegraphics[width=0.9\textwidth]{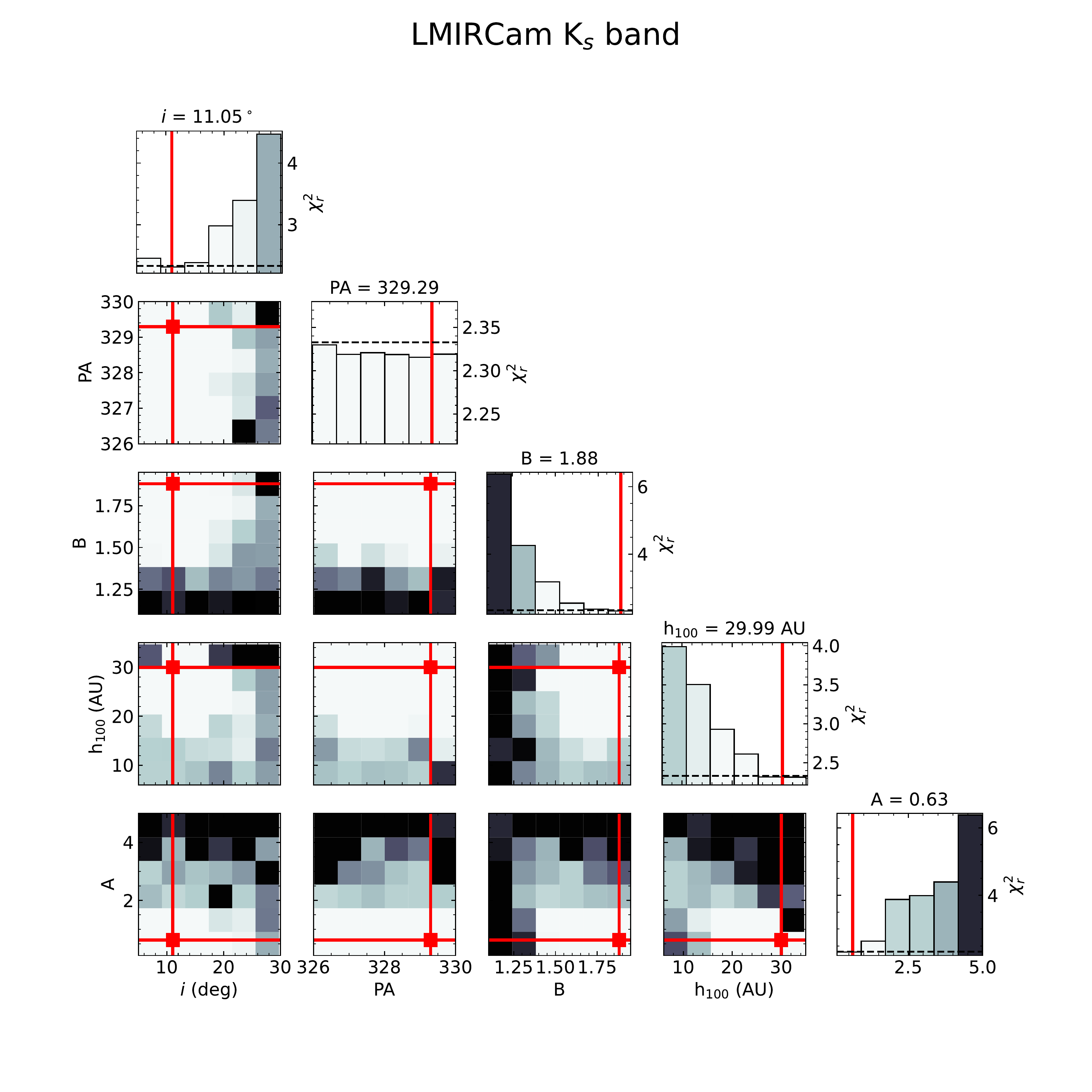}
    \caption{DE optimization of the best fit model for $K_s$ band KLIP-RDI reduction of AB Aur.  The off diagonal plots show solutions for two parameters binned 6$\times$6.  The color of each bin represents the $\chi^2_r$ of the best-fitting model within that bin.  Lighter colors indicate smaller values of $\chi^2_r$.  The diagonal plots show a one dimensional view of each parameter of the minimum $\chi^2_r$ as a function of the parameter range, and colored by $\chi^2_r$. The best fit solutions are shown by the red square/line, while the solutions within the ``acceptable range" ($\chi_r^2 \leq \chi^2_{r,min} + \sqrt{2/\nu}$) are those whose $\chi^2_r$ fall below the dashed black line in the diagonal plots.}
    \label{posteriorsK}
\end{figure}

\begin{figure}[tb]
    \centering
    \includegraphics[width=0.9\textwidth]{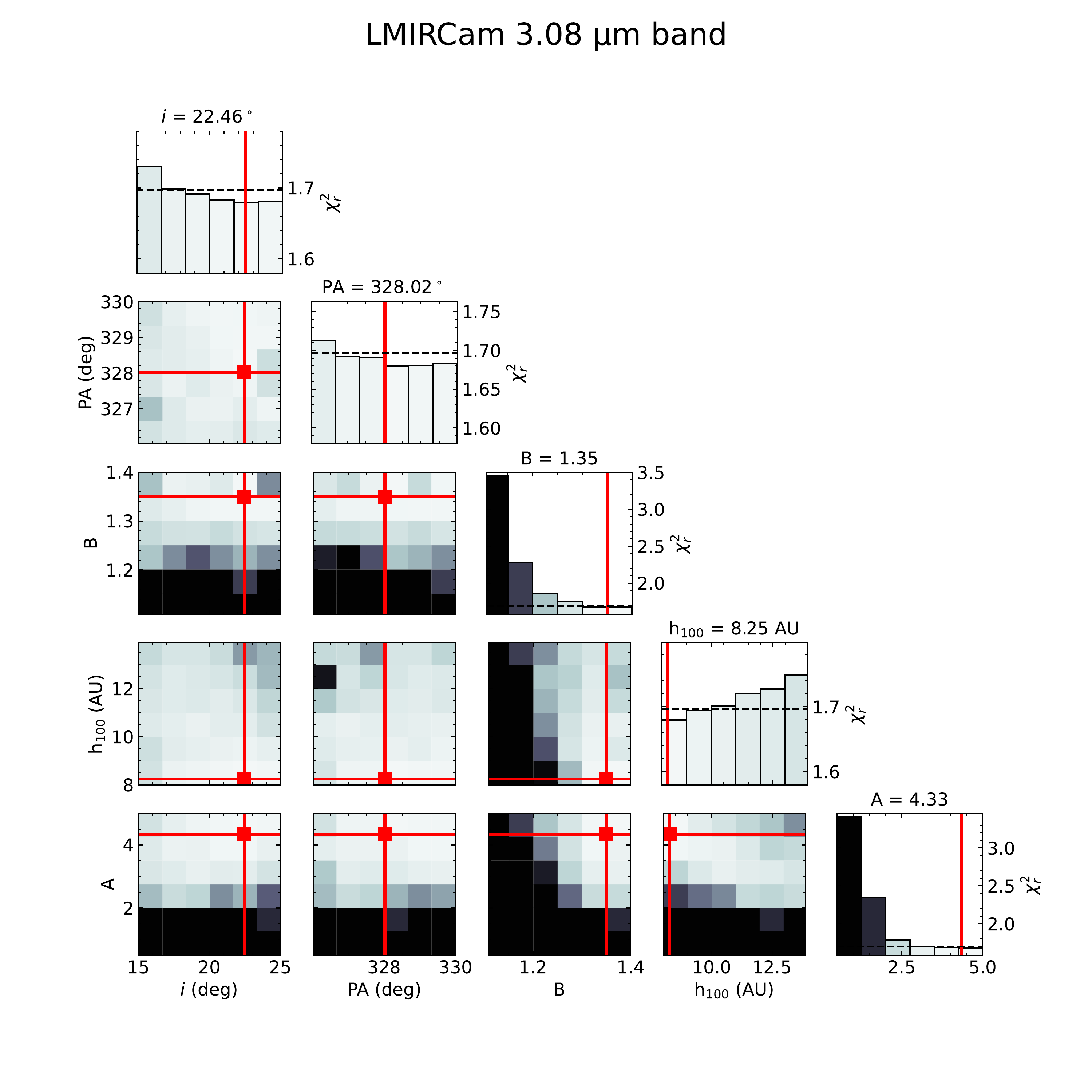}
    \caption{As Figure~\ref{posteriorsK}, but for 3.08 $\mu$m Ice band DE optimization.}
    \label{posteriorsI}
\end{figure}

\begin{figure}[tb]
    \centering
    \includegraphics[width=0.9\textwidth]{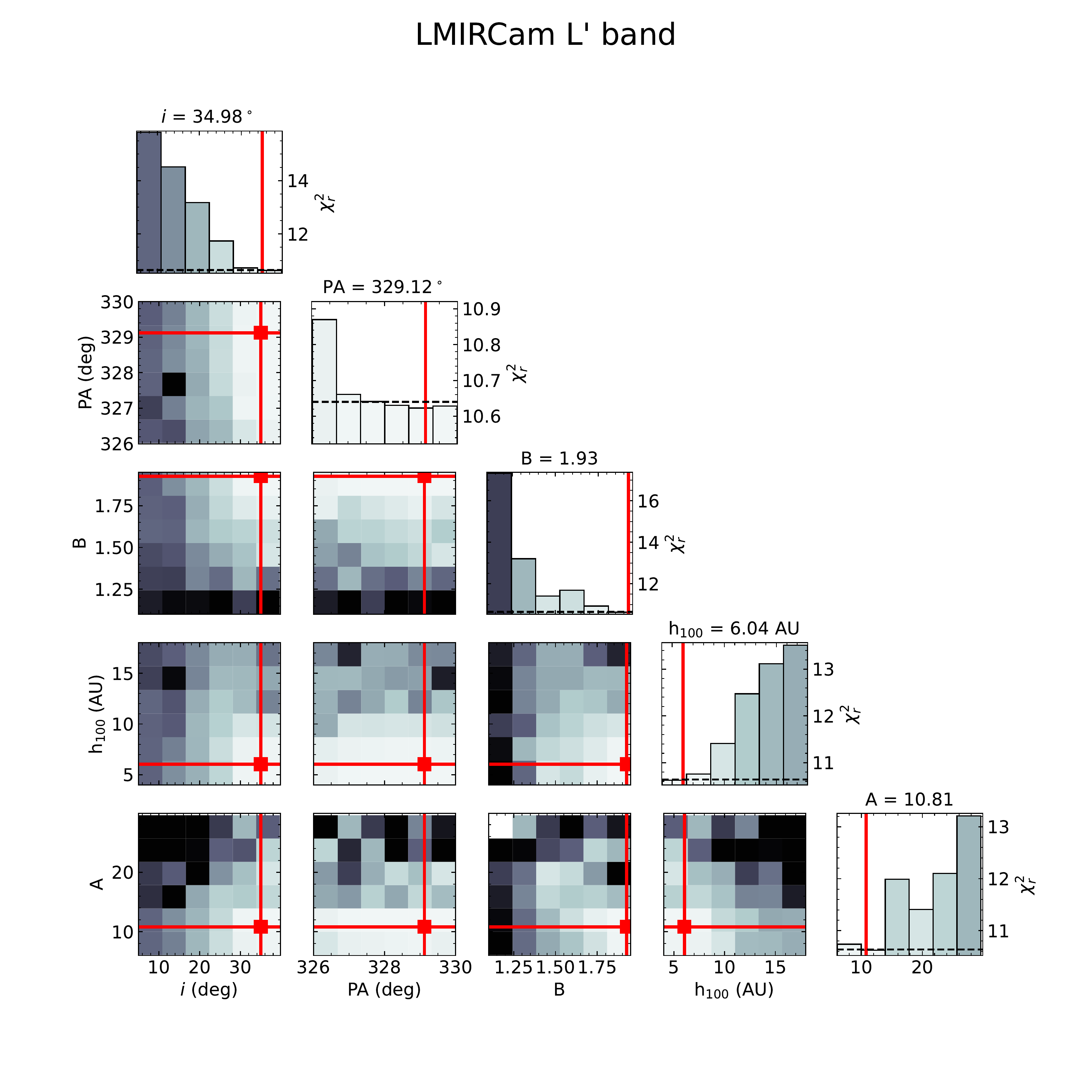}
    \caption{As Figure~\ref{posteriorsK}, but for $L'$ band DE optimization.}
    \label{posteriorsL}
\end{figure}

\begin{figure}[tb]
    \centering
    \includegraphics[width=0.85\textwidth]{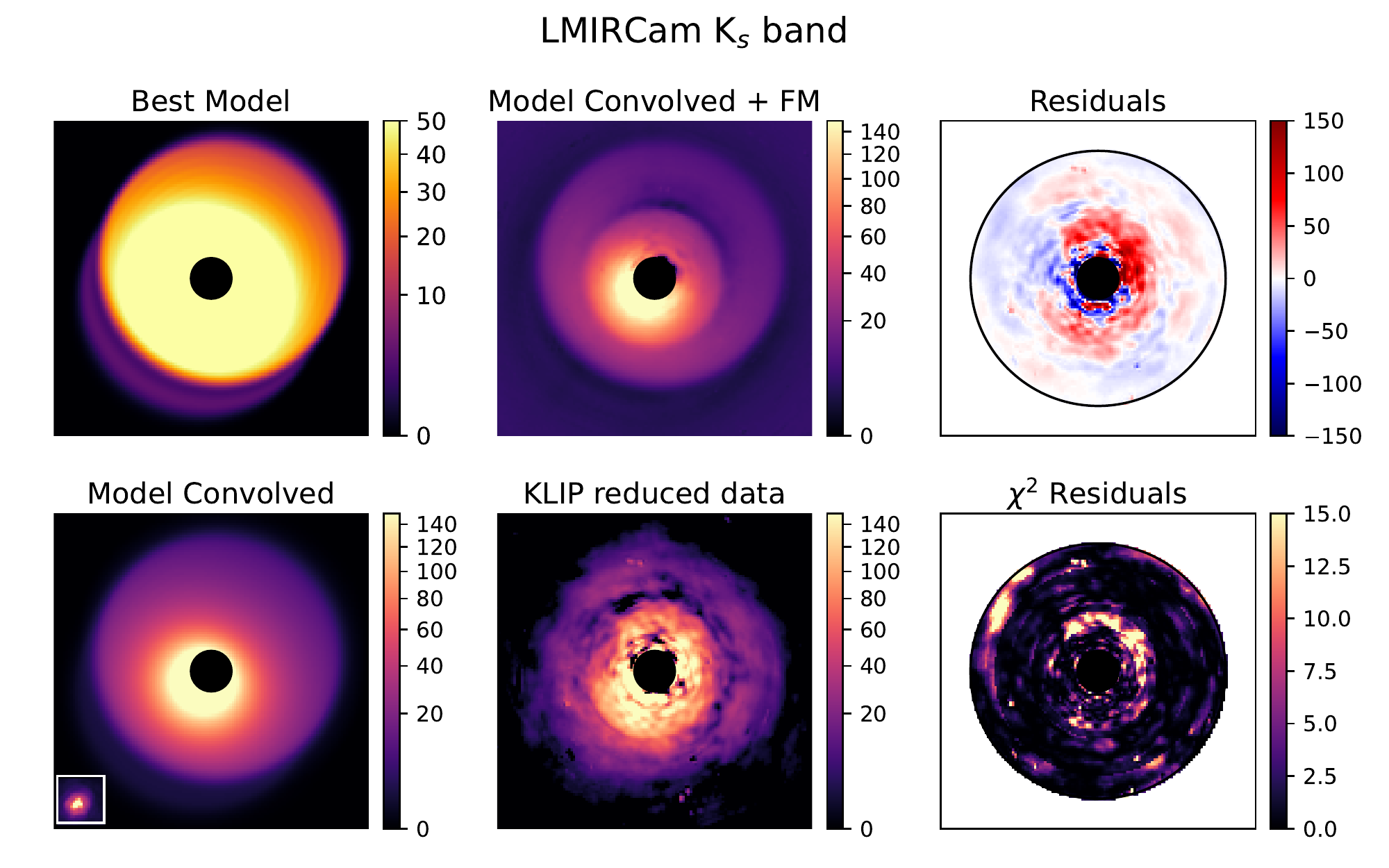}
    \caption{Best fit model and residuals from the DE optimization for $K_s$.  \textbf{top left)} Best fit model image \textbf{top center)} Best fit convolved and forward modeled image \textbf{top right)} Residuals between the model convolved + FM model and the KLIP reduced data for just the area of interest  (i.e. the disk of AB Aur). \textbf{bottom left)} Best fit model image after convolved with an observed LMIRCam PSF (shown in the offset on the bottom left of the image) \textbf{bottom center)} Observed $K_s$ band image of AB Aur after pyKLIP RDI reduction. \textbf{bottom right)} $\chi^2$ residuals between the model convolved + FM model and the KLIP reduced data for the area of interest.}
    \label{finalmodelK}
\end{figure}

\begin{figure}[tb]
    \centering
    \includegraphics[width=0.85\textwidth]{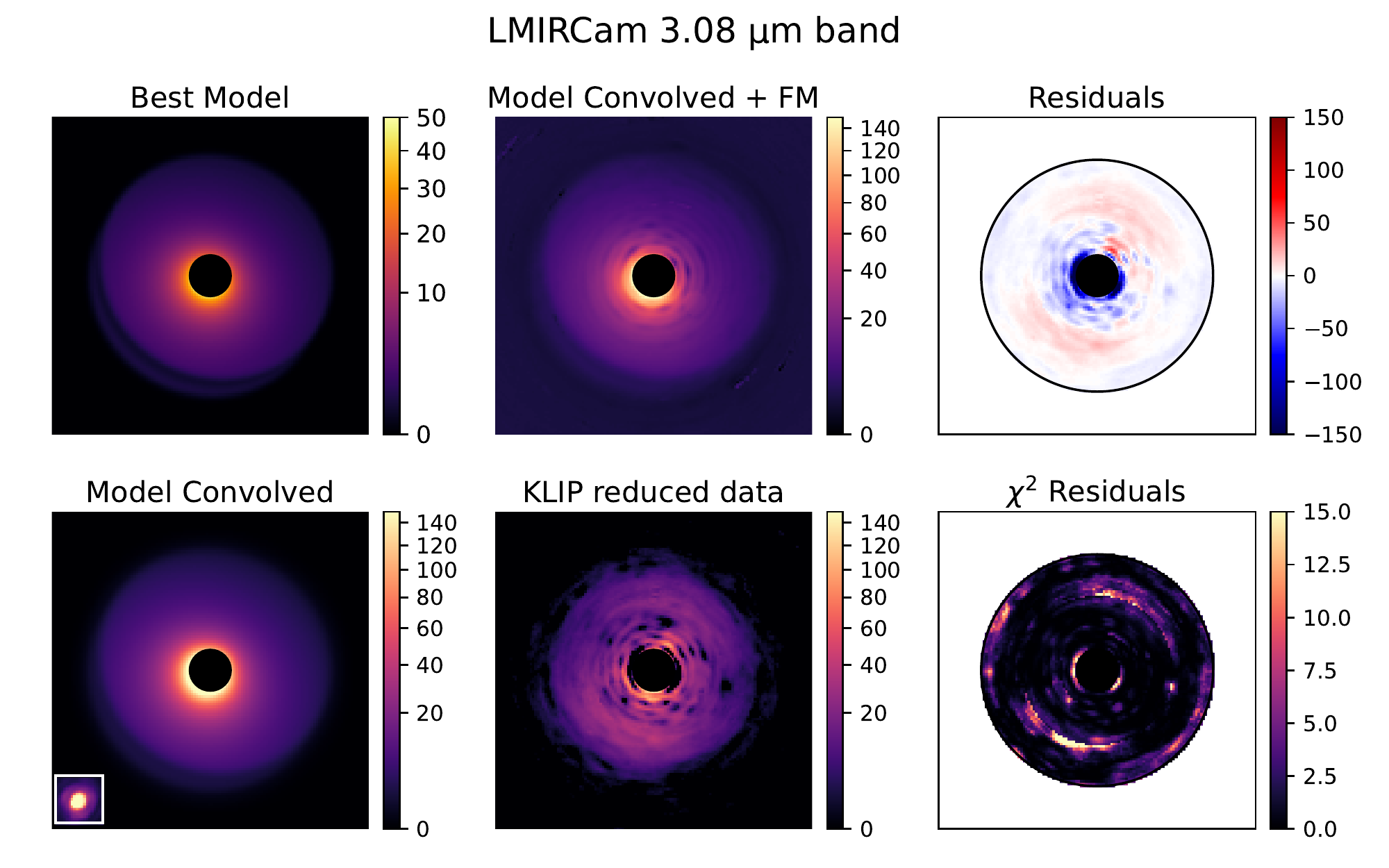}
    \caption{As Figure~\ref{finalmodelK}, but for 3.08 $\mu$m Ice band best fit model and residuals.}
    \label{finalmodelI}
\end{figure}

\begin{figure}[tb]
    \centering
    \includegraphics[width=0.85\textwidth]{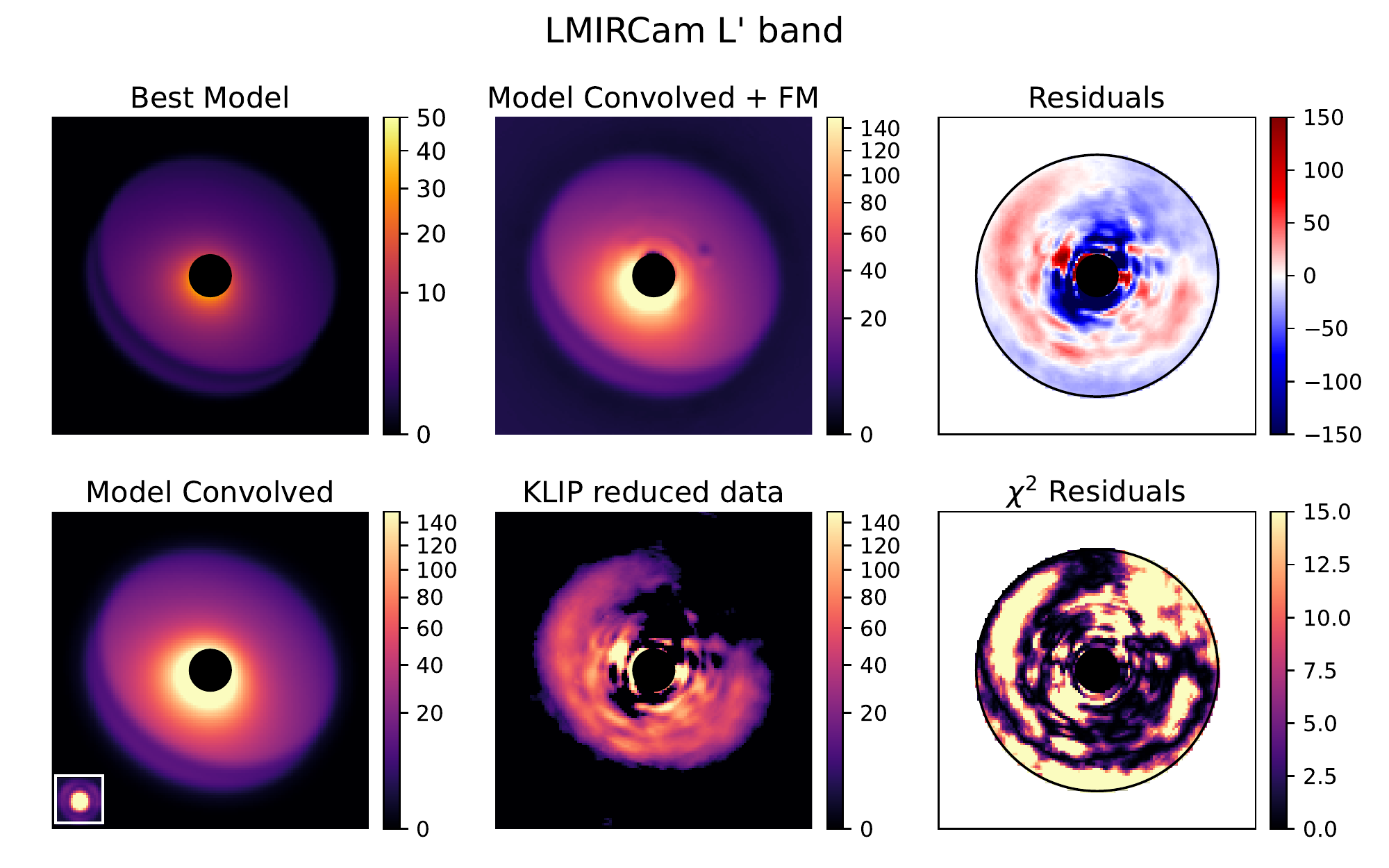}
    \caption{As Figure~\ref{finalmodelK}, but for $L'$ band best fit model and residuals.}
    \label{finalmodelL}
\end{figure}

Since the goal of this model fitting is to match the surface brightness of the disk and not to infer the real, physical disk properties, we are less concerned about finding physically-motivated or matching values between the three bands. For each band, the DE algorithm converged on a best fit solution which minimized $\chi_r^2$.  The boundaries, best fit values, and acceptable ranges for each parameter are shown in Table~\ref{tab4:finalMCFOST}.  Following \citet{Lawson2020}, models whose solutions are consistent with our observations are those with $\chi_r^2 \leq \chi_{r, min}^2 + \sqrt{2/\nu}$.  We show the minimum $\chi_r^2$, the maximum $\chi_r^2$ for acceptable solutions, and the number of models to reach convergence in Table~\ref{tab5:DEchi2}.  Figures~\ref{posteriorsK}~-~\ref{posteriorsL} shows the parameter space explored by the DE algorithm with the best fit model shown by the red cross.  The models are binned with each color showing the minimum $\chi_r^2$ found in that bin.  Figures~\ref{finalmodelK}~-~\ref{finalmodelL} shows the best fit model compared to the original data.  The residuals are only shown for the region within which $\chi^2_r$ is calculated.  Of the three bands, $L'$ had the worst convergence, most likely due to the complicated substructure in $L'$ band, namely the known spiral arms (see Jorquera et al. (2022, in press) and Figure~\ref{overlay}).  As we assume a simple, radially-symmetric disk model without taking into account the spiral features, the larger $L'$ band residuals are perhaps unsurprising. 
\begin{deluxetable}{lccccccccccc}[tb]
\centering
\tabletypesize{\scriptsize}
\tablecaption{Differential Evolution Modeling and Solutions
\label{tab4:finalMCFOST}
}
\tablehead{\colhead{Parameter} & \multicolumn{4}{c}{Parameter Boundaries}  & \multicolumn{4}{c}{Best Fit}  & \multicolumn{3}{c}{Acceptable Range} \\
\cline{2-4}
\cline{6-8}
\cline{10-12}
\colhead{} & \colhead{$K_s$} &  \colhead{3.08 $\mu$m} & \colhead{$L'$} &\colhead{} &  \colhead{$K_s$} &  \colhead{3.08 $\mu$m} & \colhead{$L'$} &\colhead{} &  \colhead{$K_s$} &  \colhead{3.08 $\mu$m} & \colhead{$L'$}
}
\startdata
$i$ (deg)   &   5$-$30  & 15$-$25 &   5$-$40 && 11.05  & 22.46  & 34.98 && 10.68$-$12.04 & 19.82$-$24.49 & 34.62$-$34.98\\
PA (deg)    &   326$-$330 & 326$-$330   &   326$-$330 && 329.29 & 328.02  & 329.12 && 326.66$-$329.79 & 326.95$-$329.95 & 328.58$-$329.9\\
$\beta$     &   1.1$-$1.95& 1.1$-$1.4  &   1.1$-$1.95 && 1.88   & 1.35 & 1.93 && 1.86$-$1.95    & 1.33$-$1.36 & 1.93$-$1.94\\
$h_{100}$ (AU) &6$-$35    & 8$-$14      &   4$-$20 && 29.99  & 8.25 & 6.04 && 29.57$-$29.99 & 8.01$-$9.06 & 6.01$-$6.05\\
$A$         &   0.1$-$5   & 0.5$-$5     &   6$-$30 && 0.63  & 4.33  & 10.81 && 0.58$-$0.66     & 3.90$-$4.70 & 10.81$-$11.22\\
\enddata
\tablecomments{PA is along the semi-minor axis, measured east of north. ``Acceptable Range" indicates parameter ranges that give $\chi_r^2 \leq \chi^2_{r,min} + \sqrt{2/\nu}$. }
\end{deluxetable} 

\begin{deluxetable}{lccc}[tb]
\centering
\tabletypesize{\scriptsize}
\tablecaption{Differential Evolution $\chi^2$ values
\label{tab5:DEchi2}
}
\tablehead{\colhead{Band} & \colhead{Minimum $\chi^2_r$} & \colhead{Acceptable $\chi^2_r$} & \colhead{\# Total Models}
}
\startdata
$K_s$ & 2.31 & 2.33 & 1519 \\
3.08 $\mu$m & 1.67 & 1.69 & 773 \\
$L'$ & 10.62 & 10.64 & 1593
\enddata
\tablecomments{Acceptable $\chi^2_r$ indicates parameter ranges that give $\chi_r^2 \leq \chi^2_{r,min} + \sqrt{2/\nu}$. }
\end{deluxetable}

\section{Throughput Variability} \label{TPuncertainty}
\begin{figure}[tb]
    \centering
    \includegraphics[width=\textwidth]{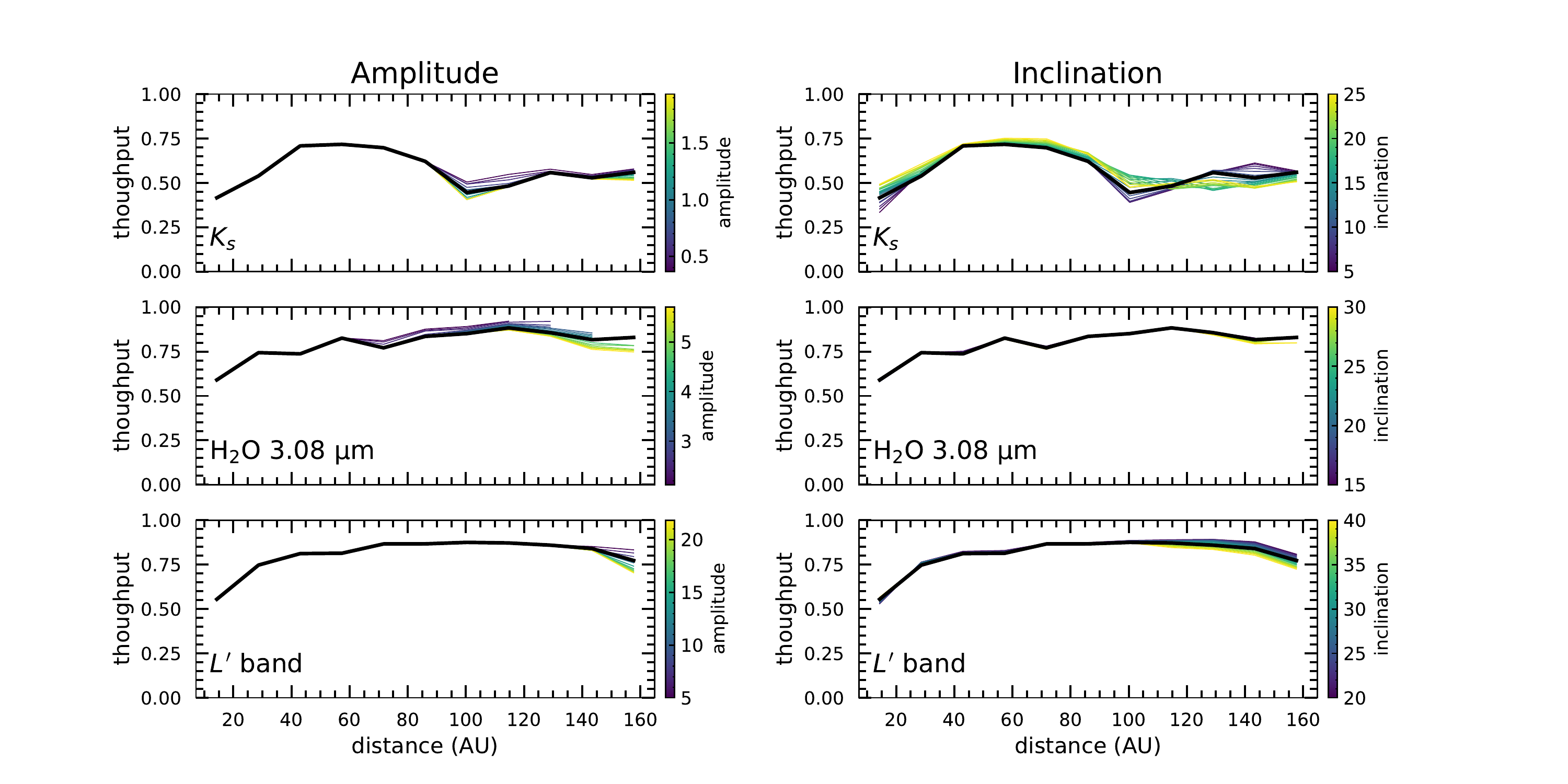}
    \caption{Azimuthally averaged throughput radial profiles for varying amplitude and inclination.  All other parameters are the fixed and best fit solutions in Tables~\ref{tab3:fixedMCFOST} and \ref{tab4:finalMCFOST}.  The black curve is the best fit profile.
    }
    \label{TP_variability}
\end{figure}

The surface brightness and disk colors are dependent on the amount of flux loss from KLIP-RDI.  Therefore, we test the variability in throughput on the brightness and geometry of the best fit model disks found in Appendix~\ref{appendix:diskmodel} in order to see how the throughput will change with varying surface brightness and inclination.  This will test the validity of using the best fit models even if the scaling factor or inclination are not perfect matches to the observed data.  Using the best fit DE solutions for each band, we vary the amplitude by a factor of 2 and the inclination within $\pm6-14^\circ$.  The model is then forward modeled using the same process outlined in Appendix~\ref{appendix:diskmodel}.  We then calculate the azimuthally averaged radial throughput.  Figure~\ref{TP_variability} shows the azimuthally averaged radial profile of the throughput for varying amplitudes and inclinations.  The black throughput curve is the best fit model throughput.  At 100 AU, where there is the large deviation in TP, it is still quite small, ranging from 0.001 in $L'$ to 0.15 in $K_s$.  
At most, the TP varies by 40\% (at 100 AU in $K_s$, at the AO control radius).  However, on average, the TP only varies by 2.5\%.  Overall, varying the amplitude and inclination affects the TP by a factor of 1.02.  Therefore, differences between the models and observed images will not cause large variations in the throughput and therefore disk color.

\end{document}